\documentclass[journal,transmag]{IEEEtran}
\usepackage{fancyhdr}

\usepackage{hyperref}
\hypersetup{
    bookmarks=true,         
    unicode=false,          
    pdftoolbar=true,        
    pdfmenubar=true,        
    pdffitwindow=false,     
    pdfstartview={FitH},    
    pdftitle={My title},    
    pdfauthor={Author},     
    pdfsubject={Subject},   
    pdfcreator={Creator},   
    pdfproducer={Producer}, 
    pdfkeywords={keyword1, key2, key3}, 
    pdfnewwindow=true,      
    colorlinks=true,       
    linkcolor=red,          
    citecolor=blue,
    filecolor=cyan,         
    urlcolor=magenta        
}

\usepackage{mathtools}
\usepackage{float}
\usepackage{caption}
\usepackage{graphicx} 
\usepackage[colorinlistoftodos]{todonotes}
\graphicspath{ {images/} }

\usepackage[linesnumbered,ruled,vlined]{algorithm2e} 

\usepackage[utf8]{inputenc}
\usepackage[english]{babel}
\usepackage{amsthm,amssymb}

\usepackage{optidef}

\theoremstyle{definition}

\DeclareMathOperator{\sgn}{sgn}

\IEEEoverridecommandlockouts                              

\title{
Polarization-adjusted Convolutional (PAC) Codes:\\ Sequential Decoding vs List Decoding}
\author{
\IEEEauthorblockN{Mohammad Rowshan, {\em Student Member, IEEE}, Andreas Burg, {\em Member, IEEE},\\ Emanuele Viterbo, {\em Fellow, IEEE}}
\thanks{M. Rowshan and E. Viterbo are with the  Department of Electrical and Computer Systems Engineering (ECSE), Monash University, Melbourne, VIC3800, Australia. E-mail: mrowshan@connect.ust.hk,  emanuele.viterbo@monash.edu. These authors' work was supported by the Australian Research Council under Discovery Project ARC DP160100528.}
 \thanks{Andreas Burg is with the Telecommunications
Circuits Laboratory (TCL), Swiss Federal Institute of Technology
(EPFL), Lausanne 1015, Switzerland E-mail: andreas.burg@epfl.ch.}
}
\begin{document}

\maketitle
\thispagestyle{empty}
\pagestyle{empty}
\thispagestyle{fancy}
\lhead{\color{blue} 
\footnotesize
Initially uploaded to arXiv on Feb 17, 2020, \url{https://doi.org/10.48550/arXiv.2002.06805}\\
Published in IEEE Trans on Vehicular Technology (Vol 70, Issue 2, Feb 2021), \url{https://doi.org/10.1109/TVT.2021.3052550}\\
Python script: \url{https://github.com/mohammad-rowshan/List-Decoder-for-Polar-Codes-and-PAC-Codes} 
or \url{https://codeocean.com/capsule/2841893}\\
Correction: page 11 (RM-Polar rate-profile, see the footnote).
}
\cfoot{}

\begin{abstract}
In the Shannon lecture at the 2019 International Symposium on Information Theory (ISIT), Ar\i kan proposed to employ a one-to-one convolutional transform as a pre-coding step before the polar transform. The resulting codes of this concatenation are called {\em polarization-adjusted convolutional (PAC) codes}. In this scheme, a pair of polar mapper and demapper as pre- and post- processing devices are deployed around a memoryless channel, which provides polarized information to an outer decoder leading to improved error correction performance of the outer code. In this paper,  the list decoding and sequential decoding (including Fano decoding and stack decoding) are first adapted for use to decode PAC codes. Then, to reduce the complexity of sequential decoding of PAC/polar codes, we propose (i) an adaptive heuristic metric, (ii) tree search constraints for backtracking to avoid exploration of unlikely sub-paths, and (iii) tree search strategies consistent with the pattern of error occurrence in polar codes. These contribute to the reduction of the average decoding time complexity from 50\% to 80\%, trading with 0.05 to 0.3 dB degradation in error correction performance within FER=$10^{-3}$ range, respectively, relative to not applying the corresponding search strategies. Additionally, as an important ingredient in Fano decoding of PAC/polar codes, an efficient computation method for the intermediate LLRs and partial sums is provided. This method is effective in backtracking and avoids storing the intermediate information or restarting the decoding process. Eventually, all three decoding algorithms are compared in terms of performance, complexity, and resource requirements.
\end{abstract}



\section{INTRODUCTION}
\label{sec:intro}
Polar codes proposed by Ar\i kan in \cite{arikan} are the first class of  channel codes with an  explicit construction that was proven to achieve the symmetric (Shannon) capacity of a binary-input discrete memoryless channel (BI-DMC) using a low-complexity successive cancellation (SC) decoder (SCD).

Polar codes are founded on the polarization effect resulting from channel synthesizing in a particular  fashion. The idea of building synthetic channels originated from the concatenated schemes \cite{arikan3} employed in the sequential decoding of convolutional codes by Massey \cite{massey} and Pinsker \cite{pinsker} in order to boost the cutoff rate. The cutoff rate is said to be "boosted" when the sum of the cutoff rates of the synthesized channels is greater than the sum of the cutoff rates of the raw channels. The key idea in boosting the cutoff rate is to build a vector channel 
where the independent copies of raw channels are transformed into multiple correlated channels.  
In Pinsker's scheme, the inner block code (with length $N$) is suggested to be chosen at random. This requires maximum likelihood (ML) decoding with prohibitive complexity. Polar codes allow using a more practical decoder with the complexity of $O(N\log N)$. 
Unlike Pinsker's scheme, where the outer convolutional transforms are identical, in multi-level coding and multi-stage decoding (MLC/MSD), originally proposed in \cite{imai} as an efficient coded-modulation technique, 
$N$ convolutional codes at different rates $\{R_i\}$ are used which consequently require a chain of $N$ outer convolutional decoders. 

On the other hand, polar coding was originally designed as a low-complexity recursive channel combining and splitting operation, where the polarization effect constrains the rates $R_i$ to either 0 or 1. 
They turned out to be so effective that no outer code was employed to achieve the original aim of boosting the cutoff rate to channel capacity. 

Nevertheless, the error correction performance of finite-length polar codes under successive cancellation (SC) decoding is not competitive, due to the existence of partially polarized channels. To address this issue, the SC list (SCL) decoding (SCLD) was proposed in \cite{tal}. This yields an error correction performance comparable to maximum-likelihood (ML) decoding. Also, it was observed that further improvement could be obtained by concatenating cyclic redundancy check (CRC) bits to polar codes. 

\begin{figure}
    \centering
    \includegraphics[width=0.8\columnwidth]{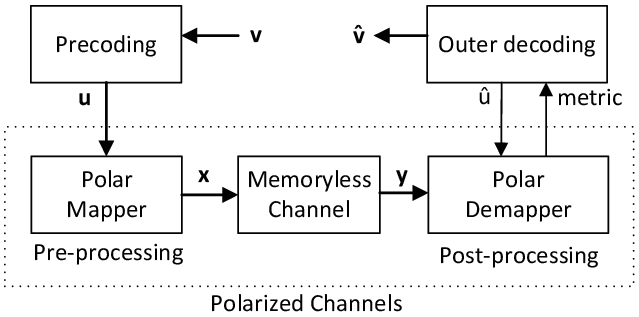}
    \caption{\vspace{-5mm} Code Concatenation} 
    \label{fig:PAC_scheme0}
\end{figure}

Recently in \cite{arikan2}, Ar\i kan proposed a concatenation of a convolutional transform with the polarization transform \cite{arikan}, inspired by the aforementioned schemes in which the message is first encoded using a convolutional transform and then transmitted over polarized synthetic channels as shown in Fig. \ref{fig:PAC_scheme0}. These codes are called ``polarization-adjusted convolutional (PAC) codes". The results show that the block error rate performance of this scheme can reach the finite-length capacity bound \cite{polyanskiy} a.k.a. dispersion bound.

Fano decoding is an efficient algorithm in terms of required hardware resources such as memory and computation resources, and it has shown a promising error correction performance. Nevertheless, Fano decoding has a high average time complexity. The motivation of this work  is to reduce the time and computational complexity at the cost of a small degradation in the practical range of frame error rate (FER), i.e. $10^{-2}$ to $10^{-4}$. Hence, this paper is concerned with the efficient implementation of Fano decoder as well as stack and list decoders for PAC codes, and compares the numerical results with classical polar codes in terms of error correction performance and complexity. The contributions of this work are given below.

\begin{itemize}
\item The Fano decoding algorithm requires  backtracking during the binary tree search. Hence, intermediate log-likelihood ratios (LLRs) and partial sums need to be updated. This update should be performed without restarting the decoding  operation or storing more than $2N-1$ intermediate LLRs and partials sums as in conventional SC decoding \cite{leroux}. In this work,  {\em partial rewinding} of the SC algorithm is proposed as an efficient approach to compute the intermediate LLRs and partial sums. 
\item The Fano metric is modified in order to improve the comparability of (unexplored) 
paths with different lengths with the current path. Furthermore, to reduce  the number of visited nodes, an adaptive bias is proposed to adjust the bias-term in the metric relative to the impact of the channel noise on the metric.  
\item A tree search strategy is proposed in which the number of diverging paths from the current best path is limited. 
This is equivalent to constraining the search to the paths in which there are a limited number of flipped bits. 
Further, this strategy is applied only to the set of bit indices called {\em critical set} where over 99\% of the errors occur. This set can reduce the time complexity by visiting fewer  nodes at the cost of negligible error-rate degradation.
\item A combination of {\em top-down} search and {\em bottom-up} search strategies for the tree search are proposed to adapt the Fano algorithm to the pattern of error occurrence in polar codes,  resulting in the faster finding of the correct path.
\item Performance and complexity comparisons of Fano decoding  with stack decoding and list decoding for polar codes and PAC codes, with  and without CRC concatenation, for different block-lengths and rate-profiles, are provided by means of simulation. 
\end{itemize}


\textbf{Paper Outline:}
Section II introduces the notations for polar codes and convolutional codes and describes their decoding algorithms. Section III illustrates polarization-adjusted convolutional transform, and describes the decoding algorithms. In Section IV, first, an efficient method for calculating the intermediate LLRs and partial sums required through backtracking in Fano decoding is proposed. Then, a heuristic path metric for Fano decoding is introduced. In Section V, strategies are described to improve Fano decoding including adaptive path metric, tree search strategies and search constraints. In Section\,VI, the distance properties of PAC codes and polar codes are compared and the implementation results are shown. Finally Section VII provides some concluding remarks.

\section{PRELIMINARIES}\label{sec:prelim}
Polarization-adjusted convolutional codes are convolutional pre-transformed polar codes. The pre-transformation (a.k.a pre-coding) is performed by a rate-1 convolutional encoding as shown in Fig. \ref{fig:PAC_scheme0}. Hence, in the following sections, we first review polar codes and convolutional codes as standalone codes, then we focus on PAC codes.

\subsection{Polar Codes and List Decoding}\label{sec:polar_codes}
A polar code of length $N = 2^n$ with $K$ information bits is denoted by $P(N,K,\mathcal{A})$, where $\mathcal{A}$ is the data index set. The information bits $d$ of length $K$ is embedded in the vector $u$ such that $u_{\mathcal{A}} = d$, and
$u_{\mathcal{A}^c} = 0$ which are called  frozen bits. The set $\mathcal{A}$ constitutes of indices of reliable sub-channels of the polarized vector channel.

A polar code is encoded as $\mathbf{x}=\mathbf{uP}_n$, where $\mathbf{P}_n=\mathbf{P}^{\otimes n}$ is the {\em polar\,transform}  defined as  the $n$-th Kronecker power of  
$\mathbf{P} \overset{\Delta}{=}{\footnotesize \begin{bmatrix}
1 & 0 \\
1 & 1
\end{bmatrix} }$.
Let $\mathbf{y}\!=\!(y_0, y_1, . . . , y_{N-1} )$ denote the output vector of a noisy channel.





\begin{figure}
    \centering
    \includegraphics[width=0.8\columnwidth]{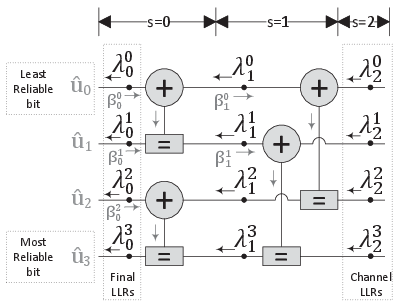}
    \caption{\vspace{-5mm}Successive cancellation factor graph for $N=4$} 
    \label{fig:factor_graph}
    \vspace{-5pt}
\end{figure}

The standard decoding method for polar codes is successive cancellation (SC) decoding in which the non-frozen bits are estimated successively based on their evolved log-likelihood ratio (LLR), denoted by $\lambda^i_s$, where $s$ is the stage of the factor graph in Fig. \ref{fig:factor_graph}. However, successive hard decisions make the SC solution sub-optimal. When decoding the $i$-th bit, if $i \notin \mathcal{A}$, 
$\hat{u}_i=0$, as $u_i$ is a frozen bit. Otherwise, the bit $u_i$ is decided by a local maximum likelihood (ML) rule 
$h(\lambda^i_0)$ in (\ref{eq:sc_hard_decision}), which depends on the estimation of previous bits, i.e., $\hat{u}_{0,{i-1}}=\hat{u}_0, ..., \hat{u}_{i-1}$
\begin{equation}
\label{eq:sc_hard_decision}
\small
\hat{u}_i = h(\lambda^i_0)= \begin{dcases*}
        0 & $\lambda^i_0 = \ln \frac{P(Y,\hat{u}_{0,{i-1}}|\hat{u}_i=0)}{P(Y,\hat{u}_{0,{i-1}}|\hat{u}_i=1)}>0$\\
        1 & otherwise\\
\end{dcases*}
\end{equation}

\normalsize




To resolve the problem of potentially erroneous decisions in the SC decoding, we can introduce the notion of a binary decision tree. In this tree, each branch on level $i$ corresponds to a decision for $u_i=1$ or $u_i=0$ and a path from the root to a leaf corresponds to a decoded codeword. 
SC decoding explores only a single path of this tree. 
Let us denote the {\em SC path} as the path in the tree obtained by following the branches with the larger likelihood (these branches are called {\em good branches} and the alternative ones are called {\em bad branches} throughout this paper) in each {\em decoding step}. A decoding step is defined as the process required to decode a bit. 
By exploring multiple or even all paths, erroneous preliminary decisions can be corrected to potentially reach ML performance. 

Since exhaustive exploration of the tree is prohibitively complex, SC list (SCL) decoding \cite{tal} performs a constrained breadth-first search (proceeding from the root to the leaves) 
which tracks only up to $L$ parallel paths, that are deemed to be the most reliable ones based on the local decisions. 

Let $\hat{u}_i[l]$ denote the estimate of $u_i$ in the $l$-th path, where $l\!\in\!\{1, 2, \ldots ,L\}$. In \cite{balatsoukas}, unlike \cite{tal}, a path metric (\mbox{\em PM}) based on LLRs magnitudes is used to measure the reliability of each path to make local decisions about which path to keep and which one to drop. The \mbox{\em PM} at $\hat{u}_i[l]$ 
is approximated by 

\begin{equation}
\label{eq:pm_func} 
PM^{(i)}_l \! = \!
\begin{dcases*}
PM^{(i-1)}_l + |\lambda^i_0[l]| & if $\hat{u}_i[l]\!\neq\!h(\lambda^i_0[l])$ \\
PM^{(i-1)}_l & otherwise	\\
\end{dcases*}
\end{equation}
\normalsize
where $PM^{(-1)}_l = 0$.

As (\ref{eq:pm_func}) shows, the path corresponding to the less likely bit value is penalized by $|\lambda^i_0|$ of that bit. 
The $L$ paths with the smallest path metrics are chosen from $2L$ paths at each step (level of the tree) and are stored in ascending order from $PM^{(i)}_1$ to $PM^{(i)}_L$.
At $N$-th step, the path with the smallest path metric $PM^{(N)}_1$ is selected as the estimated codeword.

Additionally, to compensate for the known poor distance properties of polar codes, an $r$-bit CRC is often appended to the message as an outer code to assist the decoder in error detection and finding the correct path among the $L$ paths in the list. 
However, this concatenation increases the polar code rate to $(K+r)/N$ causing a small performance degradation in the low SNR regime.  

\subsection{Convolutional Codes and Fano Decoding}\label{sec:conv_codes}
Convolutional codes (CCs) 
are a class of linear codes described by a tuple  $(n_0,k,m)$, where $k$ is the number of information bits shifted into the encoder at each time slot (usually $k = 1$), $n_0$ is the number of corresponding outputted coded bits, and $m$ is the number of previous input bits stored in a  shift-register (a.k.a. {\em memory size}) \cite{moon}. Unlike the 1-to-1 convolutional transform in PAC codes where $n_0=k=1$ as illustrated in Section \ref{ssec:encoding}, the code rate of  convolutional codes is given by $k/n_0$. 
The value $m+1$, named {\em constraint length}, determines the number of previous input bits plus the current bit that influence each coded bit. 
A larger constraint length generally provides greater resilience to bit errors. 

The relation between the input bits $d_{i-m,i}$ and one of the $n_0$ output bits $x_i$, at time-step $i$, is  obtained as a binary convolution $x_i = \sum_{j=0}^m g_j d_{i-j}$, where  $g_i\in\{0,1\}$. By representing bit sequences as polynomials in the delay variable $D$ representing a time-step in the encoder, an output sequence $x(D)$ is obtained  as $g(D) d(D)$, where  $g(D) = \sum_{j=0}^m g_j D^j$ is the {\em generator polynomial}. Different generator polynomials are used for $n_0$ outputs, only one polynomial is employed in the pre-transformation of PAC codes. 




Convolutional codes are decoded using the trellis-based Viterbi algorithm and the tree  search sequential decoding algorithms. The Viterbi algorithm is a maximum likelihood decoding method that examines the entire state space of the encoder at each step. 
We have studied Viterbi decoding of PAC codes in \cite{rowshan-lva}.  

On the other hand, the complexity of the sequential decoding is essentially independent of the  memory of the encoder, since only one encoder state is examined at each step. 
The fundamental idea behind sequential decoding is to explore only the most promising path(s). If a path to a node looks ``bad" we can discard all the paths through this node without a significant loss in the error correction performance compared to that of a maximum likelihood decoder \cite{moon}. 

In this work, we focus on Fano decoding which is a memory-efficient type of Sequential decoding algorithm. 
The {\em Fano algorithm} is a depth-first tree search, in which the decoder moves from a node  either back to its parent node or to one of its children. The Fano decoder can visit a node only if its Fano path metric $\mu_F$ is larger than or equal to a certain value called threshold $T$. Threshold takes only discrete values $0, \pm \Delta,  \pm2\Delta, \ldots$. 

Comparing the above described Fano decoding to the SC and SC list decoding described in Section \ref{sec:polar_codes}, it is instructive to note two important differences:
The SC decoding makes decisions to choose the node to visit at each step based on the branch metric. Thus, only one path is explored and the rest are discarded. 
However, SC list decoding explores multiple paths, but in contrast to Fano decoding, all of them have the same length. 
Thus, no backtracking is performed neither in SC nor in SC list decoding. Hence, only the path metric used to measure the likelihood of the paths in the sequential decoding must consider the difference in the lengths of partial paths by adding a bias, while in the SC and SC list decoding, the bias term is not required. A simpler  sequential decoding algorithm is the {\em stack decoding} \cite{moon} where the algorithm keeps a stack of size/depth $D$ of partial paths sorted with respect to the path metric. The algorithm extends the path with the best metric at the top of the stack. The stack decoding is a memory intensive algorithm with a variable time complexity that instead of backtracking as in the Fano decoding, it selects to extend the best partial path in the stack at each time step.

The metric used in the sequential decoding of convolutional codes is a probabilistic path metric. We consider the set $\mathcal{X}=\{\mathbf{a}^{(1)}, \mathbf{a}^{(2)},..., \mathbf{a}^{(M)}\}$ of $M$ partial sequences, representing partially explored paths with different lengths, to be compared.  Let $n_{max}=\max\{n_1, n_2,..., n_M\}$ denote the length of the longest sequence,  and $\tilde{\mathbf{r}}$  the partial received sequence of length $n_{max}$ symbols where each encoded symbol takes $n$ bits corresponding to $k$ information/uncoded bits, $R=k/n$. Hence, the sequences $\tilde{\mathbf{r}}$ and $\mathbf{a}^{(\ell)}$ are
\[
\tilde{\mathbf{r}}=(\mathbf{r}_0 \mathbf{r}_1 \ldots \mathbf{r}_{n_{max}-1})=
(r_0 \ldots r_{n-1} \ldots r_{nn_{max}-1})
\]
\[
\mathbf{a}^{(\ell)}=(\mathbf{a}_0^{(\ell)} \mathbf{a}_1^{(\ell)} \ldots \mathbf{a}_{n_{\ell}-1}^{(\ell)})=(a_0^{(\ell)}\ldots a_{n-1}^{(\ell)} \ldots a_{nn_{\ell}-1}^{(\ell)})
\]
Among the sequences in $\mathcal{X}$, we choose the partial sequence $\mathbf{a}^{(\ell)}$ that maximizes the a-posterior probability $P(\mathbf{a}^{(\ell)}|\tilde{\mathbf{r}})$. 
According to Bayes' rule
\begin{equation} 
\label{eq:fano_metric0}
P(\mathbf{a}^{(\ell)}|\tilde{\mathbf{r}})=\frac{P(\mathbf{a}^{(\ell)})P(\tilde{\bf r}|\mathbf{a}^{(\ell)})}{P(\tilde{\bf r})}
\end{equation}
Assuming the channels are memoryless, since the length of the sequence $\mathbf{a}^{(\ell)}$ is $n_\ell\leq n_{max}$, and there is no associated symbols in this sequence for $r_{n_{\ell}},...,r_{n_{max}-1}$, then we have 
\begin{equation}
\label{eq:fano_metric01}
P(\tilde{\mathbf{r}}|\mathbf{a}^{(\ell)})=
\prod_{j=0}^{n_{\ell}-1}P(\mathbf{r}_j|\mathbf{a}_j^{(\ell)})\prod_{j=n_{\ell}}^{n_{max}-1}P(\mathbf{r}_j)
\end{equation}
Also, we can rewrite the denominator of (\ref{eq:fano_metric0}) as 
\begin{equation}
\label{eq:fano_metric02}
P(\tilde{\bf r})=\prod_{j=0}^{n_{\ell}-1}P(\mathbf{r}_j)\prod_{j=n_{\ell}}^{n_{max}-1}P(\mathbf{r}_j)
\end{equation}
Now, by substituting (\ref{eq:fano_metric01}) and (\ref{eq:fano_metric02}) in (\ref{eq:fano_metric0}) and cancelling the common term $\prod_{j=n_{\ell}}^{n_{max}-1}P(\mathbf{r}_j)$, we have 
\begin{equation} 
\label{eq:fano_metric03}
P(\mathbf{a}^{\ell}|\tilde{\mathbf{r}})=P(\mathbf{a}^{\ell})\prod_{i=0}^{nn_{\ell}-1}\frac{P(r_i|a_i^{(\ell)})}{P(r_i)}
\end{equation}

Suppose each encoded bit occurs with equal probability, then each sequence $\mathbf{a}^{(\ell)}$ occurs with probability
 $P(\mathbf{a}^{(\ell)})=(2^{-k})^{n_{\ell}}=(2^{-nR})^{n_{\ell}}=(2^{-R})^{nn_{\ell}}$. 
Thus, by taking the base-2 logarithm of (\ref{eq:fano_metric1}), we have
\begin{equation} 
\label{eq:fano_metric1}
\log P(\mathbf{a}^{(\ell)}|\tilde{\mathbf{r}})=\sum_{i=0}^{nn_\ell-1}\Big(\underbrace{\log P(r_i|a_i^{(\ell)})}_{\text{ML-metric}}\underbrace{-\log P(r_i)-R}_{\text{path-length bias}}\Big).
\end{equation}


In order to adapt (\ref{eq:fano_metric1}) for sequential (stack or Fano) decoding of polar/PAC codes, the path metric of list decoding can be used as the ML-metric term. Note that $n$ in (\ref{eq:fano_metric1}) is 1 for PAC codes as the convolutional transform is 1-to-1 resulting in $R=1$. The simplest  path-length bias in (\ref{eq:fano_metric1}) could be a fixed bias parameter as suggested in \cite{gallager}. A different bias function based on the cumulative density function (CDF) of the evolving LLRs was proposed in \cite{trifonov}. 
Further, \cite{jeong} suggested to replace the path-length bias term with $\log(1-p_{e,i})$, where $p_{e,i}$ is the error probability of $i$-th bit-channel. 

Alternatively, in the computer science literature,
the path metric of algorithm A, a graph traversal and path search algorithm, is written in the general form of \cite{sikora}
\begin{equation} 
\label{eq:metric0}
f(\mathbf{a}^{(\ell)}) = g(\mathbf{a}^{(\ell)}) + h(\mathbf{a}^{(\ell)})
\end{equation}
where the first term measures the actual cost of the $i$-th partial path as follows, 
\begin{equation} 
\label{eq:metric1}
g(\mathbf{a}^{(\ell)}) = \sum_{j=0}^{n_\ell-1}\log P(\mathbf{r}_j|\mathbf{a}_j^{(\ell)})
\end{equation}
and the second term is a heuristic estimate for the remaining cost of completing the path to its leaf with the best metric by following the corresponding (yet unknown) extension of $\mathbf{a}^{(\ell)}$. The choice of the heuristic function $h(\mathbf{a}^{(\ell)})$, determines the tradeoff between the complexity and the risk of accidentally abandoning a path that leads to the desired optimal solution. 

We propose a heuristic to estimate $h(\mathbf{a}^{(\ell)})$ for Fano decoding of polar codes and PAC codes in Section \ref{ssec:metric}. 

\section{Polarization-adjusted Codes}\label{sec:PACs}
Polarization-adjusted convolutional codes, denoted by $PAC(N,K,\mathcal{B},\mathbf{g})$, are based on the outer convolutional transform and inner polar transform. 
One may consider PAC coding as a polar coding scheme in which the inputs to the frozen bit-channels are linear combinations of previous bits obtained by  convolutional transforms. 
Thus, given that the previous bits have been estimated correctly, the decoder can still determine the value transmitted by the corresponding "bad channels". 
In the following Sections, the encoding and decoding of PAC codes are described in detail.



\subsection{PAC Encoding}\label{ssec:encoding}
The information bits $\mathbf{d}=(d_0,d_1,...,d_{K-1})$ are first mapped to a vector $\mathbf{v}=(v_0,v_1,...,v_{N-1})$ using a rate-profile. 
The rate-profile (a.k.a. code construction) is formed based on the index set $\mathcal{B}$ such that $u_{\mathcal{B}} = d$, and $u_{\mathcal{B}^c} = 0$. Note that the constraint $v_{\mathcal{B}^c}=0$ simply leads to an irregular tree code. 

After rate-profiling, the vector $\mathbf{v}$ is transformed using a convolutional generator polynomial $\mathbf{g}=[g_0,...,g_m]$ to  $u_i = \sum_{j=0}^m g_j v_{i-j}$, where $g_i\in\{0,1\}$ as discussed in Section \ref{sec:conv_codes} (see subroutine {\em convTransform} in Algorithm \ref{alg:pac_encoding}). 
Equivalently, the convolutional transform (CT) can be represented in matrix form where the rows of an upper-triangular {\em generator matrix}\,$G$ are formed by shifting the vector $\mathbf{g} = [g_0,\ldots g_m]$. 
The number of rows equals the block-length. Given the generator matrix $\mathbf{G}$, we can encode the message block $\mathbf{v}$ as $\mathbf{u}=\mathbf{v}\mathbf{G}$. 
As a result of this pre-transformation, $u_i$ for $i\in\mathcal{B}^c$ are no longer fixed or known a priori (as 0's in $\mathbf{u}$) - unlike in conventional. In fact, these formerly frozen bits are acting as parity check (PC) bits \cite{zhang} or dynamic frozen bits \cite{trifonov}. 

\begin{figure}
    \centering
    \includegraphics[width=1\columnwidth]{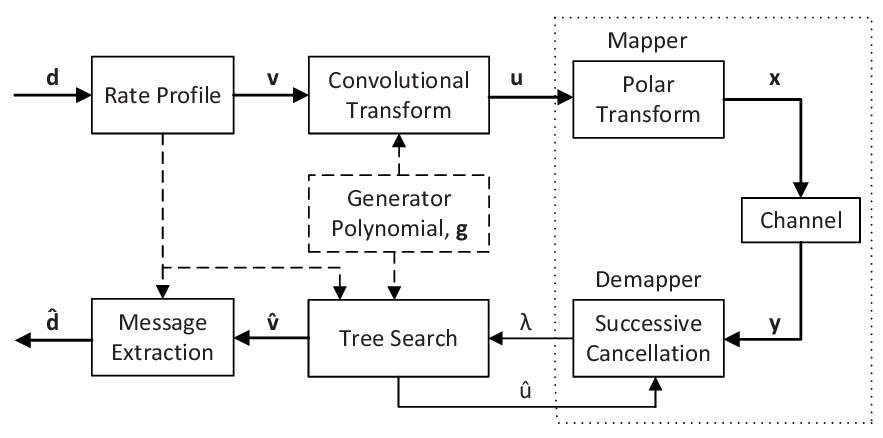}
    \caption{\vspace{-5mm} PAC coding scheme} 
    \label{fig:PAC_scheme}
    \vspace{-5pt}
\end{figure}

Then, as Fig. \ref{fig:PAC_scheme} shows, vector $\mathbf{u}$ is mapped to $\mathbf{x}$ by employing the conventional polar transform $\mathbf{P}_n$ defined in Section \ref{sec:polar_codes}. 
Hence, the the $N$-bit rate-profiled data (or block-length) should be a power of 2, i.e., $N=2^n$. In summary, the polar transformation is performed by  $\mathbf{x}=\mathbf{u}\mathbf{P}_n$. Algorithm \ref{alg:pac_encoding} summarizes the encoding process. In this algorithm, cState and currState represent the {\em current state} of the $m$-bit memory.

\begin{algorithm}
\caption{PAC Encoding}
\label{alg:pac_encoding} 
\DontPrintSemicolon
\SetKwInOut{Input}{input}
\SetKwInOut{Output}{output}
\SetKwRepeat{Repeat}{do}{while} 
\Input{ profiled information bits $\mathbf{v}$,  $\mathbf{g}$}
\Output{the codeword $\mathbf{x}$}

    $\mathbf{u} \gets$ convTrans($\mathbf{v}$, $\mathbf{g}$)\;
    $\mathbf{x} \gets$ polarTrans($\mathbf{u}$) \tcp*{Like polar encoder}
    
    \KwRet $\mathbf{x}$;

    \SetKwFunction{Fce}{convTrans}
    \SetKwProg{Fn}{subroutine}{:}{}
    \Fn{\Fce{$\mathbf{v}$, $\mathbf{g}$}}{
        cState[1,...,$|\mathbf{g}|-1$] $\gets$ [0,...,0] \tcp*{currState}
        \For{$i\gets 0$ \KwTo $|\mathbf{v}|-1$}{
            ($u_i, cState) \gets$ conv1bTrans($v_i$, cState, $\mathbf{g}$)\;
        }
        \KwRet $\mathbf{u}$;
    }

    \SetKwFunction{Fceb}{conv1bTrans}
    \SetKwProg{Fn}{subroutine}{:}{}
    \Fn{\Fceb{$v$, currState, $\mathbf{g}$}}{
        $u \gets v \cdot g_0$\;
        \For{$j\gets 1$ \KwTo $|\mathbf{g}|$}{
            \uIf{$g_j = 1$}{
                $u \gets u$ $\oplus$ currState[$j-1$]\;
            }
        }
        nextState $\gets$ [$v_i$] + currState[1,...,$|\mathbf{g}|-2$]\;
        \KwRet ($u$, nextState);
    }

\end{algorithm}

\subsection{PAC List Decoding}\label{ssec:decoding}
PAC codes as (irregular) tree codes can be decoded using the tree search algorithms discussed in Section \ref{sec:prelim}. In this section, we consider the list decoding for PAC codes which trades a fixed time complexity for a large memory requirement (to store a list of paths) and is easier to implement. Then, in the next section and the rest of the paper, we focus on Fano decoding which has a variable time complexity, but is much more memory-efficient.
Note that the list decoding in the context of convolutional codes is called {\em M-algorithm} \cite{lin3}. In the context of PAC codes, some results using list decoding were first presented by Huawei in ITW 2019 \cite{tong}. Later, we implemented list decoding for PAC codes in \cite{rowshan5} independently of \cite{yao}.

Algorithm \ref{alg:list-decoding} illustrates the list decoding approach. In the beginning, there is a single path in the list. When the index of the current bit is in the set $\mathcal{B}^c$, the decoder knows its value, usually $v_i=0$ and therefore it is encoded into $u_i$ based on the current memory state $currState$ and the generator polynomial $\mathbf{g}$ in line 7. Note that the subroutine {\em conv1bTrans} is identical with the one in Algorithm \ref{alg:pac_encoding}. Then, using the decision LLR $\lambda_0^i$ obtained in line 5, the corresponding path metric is calculated using subroutine $calcPM$. Eventually, the decoded value $u_i$ is fed back into SC process in line 9 to calculate  partial sums. 
On the other hand, if the index of the current bit is in the set $\mathcal{B}$ (see lines 19-26), there are two options for the value of $v_i$, 0 and 1, to be considered in line 24. For each option of 0 and 1, the aforementioned process for $i\in \mathcal{B}^c$ including convolutional encoding, and calculating path metric is performed and then the two encoded values $u_i=0$ and $1$ are fed back into SC\,process. The subroutines {\em updateLLRs}, {\em updatePartialSums}, and {\em prunePaths} in Algorithm \ref{alg:fano_coding} are identical to the ones used in Sc decoding and SCL decoding of polar codes.  Note that the vectors $\mathbf{\lambda}$ and $\mathbf{\beta}$ as shown in Fig. \ref{fig:factor_graph} are the LLRs and partial sums, respectively.

One can notice that the process of list decoding for PAC codes is similar to that for polar codes except for the additional convolutional re-encoding at each decoding step for which the next memory state is stored for each path. For medium and long block-lengths, we can also append CRC-bits or parity check (PC) bits to the information bits to help in detecting the correct path. To reduce the computational complexity and performance of list decoding, the methods proposed in the literature such as in \cite{rowshan3, rowshan} can be applied to PAC list decoding as well. 

List decoding, with its non-backtracking tree search approach, requires very large list sizes (typically $L=256$ or more) to reach the dispersion bound \cite{polyanskiy}, as it will be shown in Section \ref{sec:results}. More memory-efficient backtracking search algorithms such as the Fano algorithm can approach the dispersion bound at the cost of a higher average time complexity at low SNR regimes.

\begin{algorithm}[t]
\caption{List Decoding of PAC codes}
\label{alg:list-decoding}
\DontPrintSemicolon
\SetKwInOut{Input}{input}
\SetKwInOut{Output}{output}
\SetKwRepeat{Repeat}{do}{while} 
\SetKwFunction{func}{Subroutine}
\Input{channel LLRs $\lambda_n^{0,N-1}$, $\mathcal{B}$, $L$, $\mathbf{g}$}
\Output{recovered message bits $\mathbf{\hat{d}}$}
    $\mathcal{L} \gets \{1\}$ \tcp*{a single path in the list}
    $[\mathbf{\lambda}, \mathbf{\beta}]\gets$ [$\mathbf{\lambda}_n^{0,N-1}$+\{0\}, \{0\}]\;
    \For{$i\gets 0$ \KwTo $N-1$}{
        \uIf{$i \notin \mathcal{B}$}{
            \For{$l\gets 1$ \KwTo $|\mathcal{L}|$}{
                $\lambda_0^i[l]\gets $ updateLLRs($l$, $i$, $\mathbf{\lambda}[l]$, $\mathbf{\beta}[l]$)\;
                $\hat{v}_i[l]\gets 0$\; 
                {\small [$\hat{u}_i[l]$, cState[$l$]]$\gets $ conv1bTrans($v_i$, cState[$l$], $\mathbf{g}$)\;}
                $PM^{(i)}_l \gets$ calcPM($PM^{(i-1)}_l$,  $\lambda_0^i[l]$, $\hat{u}_i[l]$)\;
                $\mathbf{\beta}[l] \gets$ updatePartialSums($\hat{u}_i[l]$, $\mathbf{\beta}[l]$)\;
                
            }
            
        }
        \Else{
            \For{$l\gets 1$ \KwTo $|\mathcal{L}|$}{
                $\mathcal{L} \gets$ duplicatePath($\mathcal{L}$, $l$, $i$, $\mathbf{g}$)\;
            }
                \uIf{$|\mathcal{L}| > L$}{
                    $\mathcal{L} \gets$ prunePaths($\mathcal{L}$)\tcp*{\footnotesize like SCLD} 
                }

        }
    }
    $\mathbf{\hat{d}} \gets$ extractData($\hat{v}_1^N[0]$)\;
    \KwRet $\mathbf{\hat{d}}$;

    \SetKwFunction{FdL}{duplicatePath}
    \SetKwProg{Fn}{subroutine}{:}{}
    \Fn{\FdL{$\mathcal{L}$, $l$, $i$, $\mathbf{g}$}}{
        $\mathcal{L}\gets\mathcal{L}\cup \{l^\prime\}$ \tcp*{{\footnotesize path $l^\prime$ is a copy of path $l$}}
        $\lambda_0^i[l]\gets$ updateLLRs($l$, $i$, $\mathbf{\lambda}[l]$, $\mathbf{\beta}[l]$)\;
        ($\hat{v}_i[l]$, $\hat{v}_i[l^\prime]$) $\gets$ (0, 1)\;
        [$\hat{u}_i[l]$, cState[$l$]] $\gets$ conv1bTrans($\hat{v}_i[l]$, cState[$l$], $\mathbf{g}$)\;
        [$\hat{u}_i[l^\prime]$, cState[$l^\prime$]]$\gets$  conv1bTrans($\hat{v}_i[l^\prime]$, cState[$l$], $\mathbf{g}$)\;
        $PM^{(i)}_l \gets$ calcPM($PM^{(i-1)}_l$,  $\lambda_0^i[l]$, $\hat{u}_i[l]$)\;
        $PM^{(i)}_{l^\prime} \gets$ calcPM($PM^{(i-1)}_{l}$,  $\lambda_0^i[l]$, $\hat{u}_i[l^\prime]$)\;
        $\mathbf{\beta}[l] \gets$ updatePartialSums($\hat{u}_i[l]$, $\mathbf{\beta}[l]$)\;
        $\mathbf{\beta}[l^\prime] \gets$ updatePartialSums($\hat{u}_i[l^\prime]$, $\mathbf{\beta}[l]$)\;
        \KwRet $\mathcal{L}$;
    }
        
    \SetKwFunction{Fpm}{calcPM}
    \SetKwProg{Fn}{subroutine}{:}{}
    \Fn{\Fpm{$PM$, $\lambda_0$, $\hat{u}$}}{
        \uIf{$\hat{u} = \frac{1}{2}(1-\sgn(\lambda_0))$}{
            $PM=PM$\;
        }\Else{
            $PM=PM+|\lambda_0|$\;
        }
        \KwRet $PM$;
    }

\end{algorithm}

\section{Fano Decoding of PAC Codes}\label{sssec:fano-decoding}


In this section, we first briefly explain the fundamentals of the Fano algorithm detailed in Algorithm \ref{alg:fano_coding} followed by the details of the proposed algorithm for updating the intermediate information required in the SC decoding process in the backtracking (Section \ref{ssec:rewind-sc}) and our novel path metric (Section \ref{ssec:metric}).
Then, we present several improvements to the Fano algorithm in order to reduce the time complexity in the following Section \ref{sec:low-complexity-fano-decoding}.

In the Fano algorithm, the decoder starts with the origin node ($i=0$) and examines a sequence of adjacent nodes. At any step corresponding to the non-frozen bits in vector $\mathbf{v}$, it either moves forward to one of the successor nodes or moves backward to the non-frozen predecessor of the current node.
The branch metric $m_i$ is correspondingly added to the current path metric $\mu_{i-1}$ during forward movement (in lines 9 \& 16-17 of Algorithm \ref{alg:fano_coding}) or restored from memory during the backward movement. The algorithm stops when it reaches a terminal node ($i=N$). The search through the code tree is guided by a threshold $T$ on the path metric (with initial value of $T=0$). If the metric becomes less than the threshold as the algorithm follows the current path (line 42 of Algorithm \ref{alg:fano_coding}), the search is backed up and another path is followed (Algorithm \ref{alg:move_back} is called in line 58 of Algorithm \ref{alg:fano_coding}). If no paths can be found with a metric above the threshold, the threshold value is lowered (in line 26 of Algorithm \ref{alg:fano_coding}) and the process is continued. A node in the tree may be visited more than once in the forward direction but a lower threshold each time. 
The algorithm eventually reaches a terminal node and stops. For more details on the Fano algorithm, see \cite{moon}. 

Note that ({\em i}) the Fano algorithm proposed here stores the path metric of good branch (the one with larger metric) and bad branch (the one with smaller metric) as well as memory states along the current path, ({\em ii}) the subroutines {\em updateLLRs} and {\em updatePartialSums} in Algorithm \ref{alg:fano_coding} and the rest of the paper are identical to the ones used in SC decoding of polar codes, ({\em iii}) Algorithm \ref{alg:move_back} is called in line 44 to find a bit index  that satisfies the threshold in order to move back, and ({\em iv}) {\em toDiverge} indicates that the branch with smaller metric should be chosen at information bit $j$ and this choice is flagged in the $j$-element of vector $\delta$. The rest of the Algorithm  \ref{alg:fano_coding} and other algorithms are explained and referred to in the rest of the paper.

\subsection{Partial Rewind of SC Algorithm}\label{ssec:rewind-sc}

The Fano algorithm performs forward and backward traversals in the decoding tree: while in the forward traversal, the calculation of the required intermediate LLRs and partial sums is straightforward and linear, a more sophisticated approach is required for the backward traversal or partial rewind of SC algorithm.
Suppose that we want to move back from the $i_{curr}$-th bit to the $i_{start}$-th bit: 
First, we need to re-calculate $\lambda_0^{i_{start}}$ and a number of intermediate LLRs. Since these LLRs are updated in-place when computing the metrics in natural order, they may no longer be available. 
As explained in \cite{leroux}, efficient decoders store at most $N-1$ intermediate/decision LLRs for decoding bits $0$ to $N-1$, of which $N/2^{n-s}$ are associated to stage $s$ ($0\leq s\leq n-1$) of the LLR calculation algorithm.  The number of intermediate LLRs to be updated varies between one, when moving from a bit with odd index $i_{curr}$ to $i_{curr}-1$, and in extreme cases $N-1$, when moving from $i_{curr}\geq N/2$ to $i_{start}<N/2$. 

In general, 
up to $\log_2 N$ stages should be activated to calculate the decision LLR at bit $i_{start}$, $\lambda_0^{i_{start}}$. 
The first stage to be activated (from right to left in Fig. \ref{fig:factor_graph}) is determined by {\em find first set} (ffs) operation, here, $set$ means 1, on the binary representation of bit index $x$, i.e., bin$(x)=x_{n-1}...x_1x_0$. The modified version of ffs \cite{leroux} is defined below. Note that we assume the decoding is performed in natural order.

\begin{equation}
\label{eq:ffs}
\small
\text{ffs}^*(x_{n-1}...x_1x_0)= \begin{dcases*}
        \min(j):x_j=1 & $x>0$,\\
        n-1 & $x=0$\\
\end{dcases*}
\end{equation}

When $i_{curr}$, the index of the current bit, is odd, $\text{ffs}^*(\text{bin}(i_{curr}))=0$, and $i_{start}=i_{curr}-1$, we can calculate the decision LLR, $\lambda_0^{i_{start}}$, directly according to the $f$-node operation without any need to update the intermediate LLRs.  As a consequence, when moving back to bit index $i_{start}<i_{curr}-1$, we need to consider the ffs$^*$ of $i_{curr}-1$ and/or $i_{start}-1$ if $i_{curr}$ and $i_{start}$ both or either one is odd. This is controlled in lines 1-4 of Algorithm \ref{alg:inter-LLR}.  
Note that the stages to be updated are not necessarily $s=\text{ffs}^*(\text{bin}(i_{start})),...,1,0$, but the deepest stage to be updated,  $s_{max}$, is

\begin{equation}
\label{eq:smax}
\small
s_{max} = \{\max(s) : s=\text{ffs}^*(\text{bin}(i_m)), i_{start}\leq i_m\leq i_{curr} \}
\end{equation}

The relation (\ref{eq:smax}) finds the deepest stage in the factor graph (from left to right in Fig. \ref{fig:factor_graph}) at which  the LLRs have been updated/overwritten while decoding bit $i_{start}$ to $i_{curr}$. 
If $s_{max} \geq \text{ffs}^*(\text{bin}(i_{start}))$, we need to move back further to the bit $i_{-1}$ at which $s_{max} = \text{ffs}^*(\text{bin}(i_{-1}))$. 
The subroutine $findsMaxPos$ in Algorithm \ref{alg:inter-LLR} performs the operation of finding $i_{-1}$. 

{\em Example:} Suppose the block-length is $N=4$ and we are decoding bit $i_{curr}=3$. The intermediate LLRs vector is [$\lambda^3_1$, $\lambda^2_1$], excluding the decision LLR, $\lambda^3_0$ (see Fig. \ref{fig:factor_graph}). Now, if we need to go one step back to bit $i_{start}=2$, since $i_{curr}=3$ is odd, we do not need to update the intermediate LLR vector, i.e., $\lambda^2_0$ can be directly calculated. However, for moving back to $i_{start}=1$, since $i_{start}$ is odd, we need to find $s_{max}=1$ and calculate [$\lambda^1_1$, $\lambda^0_1$]. Only after this update of the intermediate LLRs it is possible to calculate the decision LLR $\lambda^1_0$ and rewind the SC algorithm.

Note that the partial sums vector, $\beta$, is also updated in lines 12 and 19 during the aforementioned process.

Algorithm \ref{alg:inter-LLR} shows an efficient approach for updating the intermediate LLRs. 

\begin{algorithm}
\caption{updateLLRsPSs: Updating intermediate LLRs \& partial sums for partial rewinding}
\label{alg:inter-LLR}
\DontPrintSemicolon
\SetKwInOut{Input}{input}
\SetKwInOut{Output}{output}
\SetKwRepeat{Repeat}{do}{while} 
\Input{$i_{start}$, $i_{curr}$,  $\hat{\mathbf{u}}$,  $\mathbf{\lambda}$, $\mathbf{\beta}$}
\Output{updated $\mathbf{\lambda}$, updated $\mathbf{\beta}$}
    \uIf{$i_{curr}\%2 \neq 0$}{
        $i_{curr} \gets i_{curr} - 1$\;
    }
    \uIf{$i_{start}\%2 \neq 0$}{
        $i_{start} \gets i_{start} - 1$\;
    }
    $s_{start} =$ ffs$^*(i_{start})$ \tcp*{c.f (\ref{eq:ffs})}
    $s_{max} \gets \text{sMax}(i_{start}, i_{curr})$
        \tcp*{c.f (\ref{eq:smax})}
    \uIf{$s_{start} \leq s_{max}$}{
        $i_{-1}=$find\_sMaxPos($s_{start}$,$s_{max}$, $i_{start}$)\;
        $\mathbf{\beta}\gets$ \text{updatePSBack}($i_{-1}$, $s_{max}$, $\hat{\mathbf{u}}$)\;
        \For{$i\gets i_{-1}$ \KwTo $i_{start}$}{
            $\mathbf{\lambda}\gets$ \text{updateLLRs}($i$, $\mathbf{\lambda}$, $\mathbf{\beta}$)\;
            $\mathbf{\beta}\gets$ \text{updatePartialSums}($i$, $\hat{u}_{i}$, , $\mathbf{\beta}$)\tcp*{\footnotesize Identical w/ SCD}
        }
    }\Else{
        $\mathbf{\lambda}\gets$ \text{updateLLRs}($i_{start}$, $\mathbf{\lambda}$,  $\mathbf{\beta}$)\;
    }
    \KwRet $[\mathbf{\lambda}, \mathbf{\beta}]$;        

        


\SetKwFunction{FbetaBack}{updatePSBack}
\SetKwProg{Fn}{subroutine}{:}{}
\Fn{\FbetaBack{$i_{-1}$, $s_{max}$, $\hat{\mathbf{u}}$}}{
    $k\gets2^{s_{max}}$\;
    \For{$i\gets i_{-1}+1-k$ \KwTo $i_{-1}$}{
        $\mathbf{\beta}\gets$ \text{updatePartialSums}($i$, $\hat{\mathbf{u}}_{i}$,  $\mathbf{\beta}$)\;
    } 
    \KwRet $\mathbf{\beta}$;
}
\SetKwFunction{Fismax}{find\_sMaxPos}
\SetKwProg{Fn}{subroutine}{:}{}
\Fn{\Fismax{$s_{start}$, $s_{max}$, $i_{-1}$}}{
    $s^{\prime}\gets s_{start}$\;
    \While{$s^{\prime} < s_{max}$ }{
        $i_{-1} \gets i_{-1} - 2$\;
        \uIf{$i_{-1}>0$}{
            $s^{\prime}\gets\text{ffs}^*(i_{-1})$ \tcp*{c.f (\ref{eq:ffs})}
        }\Else{
            $s^{\prime} \gets n$\;
        }
    } 
    \KwRet $i_{-1}$;
}

\end{algorithm}

\subsection{Heuristic Path Metric}\label{ssec:metric}
The Fano path metric for each examined node plays an important role in the backtracking since it provides an indication for how likely it is that the partial path to the current node is correct. Efficient backtracking relies on this metric to a) select a point to branch off the currently best (possibly erroneous) path to explore promising alternative solutions and to b) abandon unlikely paths based on comparing their path metrics with the threshold $T$.

To provide such a metric, we follow the generic approach outlined in (6):  the first term corresponds to the metric in list decoding while the second term is used to account for the different candidate path lengths in the Fano decoding. 
For every partial sequence $a^{(\ell)}$, we define the following  metric:
\begin{equation} 
\label{eq:pac_metric0}
\begin{multlined}
\mu_\ell=M(a^{(\ell)},\mathbf{y})=\sum_{j=0}^{n_\ell-1}\log P(\hat{u}_j^{(\ell)}|\hat{\mathbf{u}}_{0,j-1}^{(\ell)},\mathbf{y})\\ + \sum_{j=n_\ell}^{N-1}\log E_{\mathbf{y}}\big[P(u_j|\mathbf{u}_{0,j-1},\mathbf{y})\big]
\end{multlined}
\end{equation}
The second term is an expected metric for the continuation of the partial path with length $N-n_i$. Based on our observation of the  actual metric obtained during decoding with or without backtracking, a good estimation of the second term, in case there is no error in the received signals, is $E_{\mathbf{y}}\big[P(u_j|\mathbf{u}_{0,j-1},\mathbf{y})\big]
\approx 1-p_{e,j}$, where $p_e$ is the error probability of the bit-channels which can be obtained from the methods used for the construction/rate-profile of polar codes. 

Let us define the {\em expected metric} $B=E_{\mathbf{y}}[\mu_{N-1}]$ for the full-length path and the expected metric of the remaining partial path as
\begin{equation}
\label{eq:B}
B=\sum_{j=0}^{N-1} \log(1-p_{e,j})
\end{equation}
\begin{equation}
\label{eq:Bcmp}
B^c_{i}=\sum_{j=i+1}^{N-1} \log(1-p_{e,j})=B-\sum_{j=0}^{i} \log(1-p_{e,j})
\end{equation}
where $\log(1-p_{e,j})$ is the {\em estimated branch metric}. 
Now, we can rewrite (\ref{eq:pac_metric0}) as a recursion as follows:

\begin{equation} 
\label{eq:pac_metric01}
\begin{multlined}
\mu_j=\mu_{j-1}+m_j-\log(1-p_{e,j})
\end{multlined}
\end{equation} 
where $m_j=\log(P(\hat{u}_j|\hat{\mathbf{u}}_{0,j-1},\mathbf{y}))$ is the actual  branch metric and $\mu_{-1}=B$. Note that since the initial metric is $\mu_{-1}=B$, at each decoding step, the actual branch metric $m_j$ is added and instead the estimated metric of the corresponding branch is deducted to maintain the relation in (\ref{eq:pac_metric0}). Hence, although (\ref{eq:pac_metric01}) looks similar to the metric in \cite{jeong}, the initial value and the foundation of the metric are quite different (in \cite{jeong}, $\mu_{-1}=0$). 
Furthermore, one can optimize the FER performance by tuning the bias term, $\log(1-p_{e,i})$. In particular, if the SNR dependent method in \cite{trifonov3} is used to obtain $p_{e,i}$, 
one can improve FER performance, by changing the design-SNR.
\section{Low-complexity Fano Decoding}\label{sec:low-complexity-fano-decoding}
In this section, we introduce an adaptive path metric depending on the noise level and different search strategies to limit the search space.
\subsection{Adaptive Path Metric}\label{ssec:ad-metric}
\begin{figure}
    \centering
    \includegraphics[width=1\columnwidth]{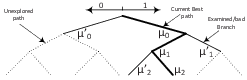}
    \caption{Decoding tree:  $\mu_j$s are the path metrics of the current best path (solid thick line) from the root to a node at level $j$ and the $\mu^\prime_j$s are the path metrics of the branches (solid thin line) diverging from the current best path.} 
    \label{fig:decoding_tree}
    \vspace{-5pt}
\end{figure}

A bit channel $i$ with low reliability contributes to the metric update depending on the noise level, i.e., $\mu_{i}$ can be significantly smaller than $\mu_{i-1}$ (due to change in the magnitude and/or sign of the decision LLR) in the presence of large channel noise. 
This impact on the path metric can accumulate over time leading to a significant deviation from the expected metric in (\ref{eq:B}). Recall that due to channel dependency, a change in the channel LLR of one channel can affect the other low-reliability bit channels as well. Consequently, the metric of  most of the examined branches denoted by $\mu^\prime$ in Fig. \ref{fig:decoding_tree} are most likely greater than the threshold, i.e., $\mu^\prime_i > T$ for $i<i_{curr}$, where $i_{curr}$ is defined in Section\,\ref{sssec:fano-decoding}. 
This causes a large delay due to the exploration of many sub-paths during backtracking. Hence, the metric estimate for the path continuation represented by the second term in (\ref{eq:pac_metric0}), 
is not in a fair way comparable with the actual metric of the current path as discussed in the previous section. 

To compensate for such deviation, we suggest adapting the estimate (\ref{eq:Bcmp}) for the continuation of partial paths relative to the impact of the channel noise on the actual metric. 
This adaptation can be realized by a scaling factor $\alpha$ for the logarithm of the probability in (\ref{eq:Bcmp}) which in effect adapts the expected probability to the noise level. The effect of this scaling is as follows: 
$\alpha\log E_{\mathbf{y}}\big[P(u_j|\mathbf{u}_{0,j-1},\mathbf{y})\big]=\log \left(E_{\mathbf{y}}\big[P(u_j|\mathbf{u}_{0,j-1},\mathbf{y})\big]\right)^\alpha$. Since $\alpha\geq 1$ and $P(u_j|\mathbf{u}_{0,j-1},\mathbf{y})<1$, then $\left(E_{\mathbf{y}}\big[P(u_j|\mathbf{u}_{0,j-1},\mathbf{y})\big]\right)^\alpha$ becomes smaller, accounting for a larger noise variance.

The value of $\alpha$ is determined after visiting the nodes of the current path to some level of decoding tree. This level should cover a sufficient number of  low-reliability bit-channels to reflect the noise effect on the metric in a fair way. Until this level/bit index denoted by $i_{bu}$ in lines 43 and 46 of Algorithm \ref{alg:fano_coding2}, we do not perform backtracking although the metric drops below the threshold, $T$ (as seen in lines 43-44 where the threshold is updated). Then, the scaling factor is obtained by

\begin{equation}
\label{eq:pac_metric00}
\alpha =  \frac{\sum_{j=0}^{n_k}\log P(\hat{u}_j^{(\ell)}|\hat{\mathbf{u}}_{0,j-1}^{(\ell)},\mathbf{y}) }{ \sum_{j=0}^{n_k}\log E_{\mathbf{y}}\big[P(u_j|\mathbf{u}_{0,j-1},\mathbf{y})\big]}
\end{equation}

This adaptation can be performed when $\alpha>1$, i.e., when the actual metric is larger than the expected metric. In practice, a quantized version of this factor is more convenient to use in fixed-point arithmetic. Hence, $\alpha_q=\lceil\frac{\alpha}{\Delta_q}\rceil\Delta_q$, where $\Delta_q$ is an integer. 
For instance, in decoding $PAC(128,64)$, we first follow the current best path to bit $i_{bu}=38$. By taking $\Delta_q=2$ and the effect of the ceiling operator, an effective value is obtained which further reduces the complexity  
with almost no degradation in performance. In low and medium code rates, one can choose to calculate $\alpha$ after the initial sequence of low-reliability bits, where the associated values in vector $v$ are 0 (equivalent to the frozen bit-channels in polar codes).

After obtaining $\alpha$, we need to update not only the metric of the current path, but also the metric of the examined branches, $\mu^\prime_j$ in Fig. \ref{fig:decoding_tree}, along the current path.

To update the computed metrics we simply add the difference between the updated bias $\alpha B_j^c$ and the initial bias $B_j^c$ to $\mu_j$ and $\mu_j^\prime$.
\begin{equation}
\label{eq:updatingMetric}
\mu^\prime_j=\mu^\prime_j + (\alpha-1) B_j^c
\end{equation}

Thus, the metrics are computed by considering $\alpha$ in the next decoding steps as 
\begin{equation} 
\label{eq:pac_metric02}
\begin{multlined}
\mu_j=\mu_{j-1}+m_j-\alpha\cdot \log(1-p_{e,j})
\end{multlined}
\end{equation}
Lines 10 and 16-17 of Algorithm \ref{alg:fano_coding} include $\alpha$ which is initialized at the beginning of the decoding, line 3 ($\alpha=1$). The calculation of $\alpha$ and the metric updating process are shown in Algorithm \ref{alg:fano_coding2}, lines 46-53.

For hardware implementation, we are interested in simple arithmetic operations. Here, we suggest using an LLR-based metric instead of the metric based on the probability. To this end, we need to define $m_j$ based on $\lambda_0^j$. 
\begin{eqnarray} 
\label{eq:pac_metric03}
m_j(\lambda_0^j,\hat{u}_j) \!\!&=&\!\!\log(P(\hat{u}_j|\hat{\mathbf{u}}_{0,j-1},\mathbf{y}))=\log\left(\frac{e^{(1-\hat{u}_j)\lambda_0^j}}{e^{\lambda_0^j}+1}\right) \nonumber \\ 
\!\!&=&\!\!\log\left(1+e^{-(1-2\hat{u}_j)\lambda_0^j}\right)^{-1}
\end{eqnarray}
where the last equality holds only for $\hat{u}_j$ = 0 and 1. 
Now, if $\hat{u}=\frac{1}{2}(1-\sgn(\lambda_0^j))$, the term $e^{-(1-2\hat{u})\lambda_0^j}=e^{-|\lambda_0^j|}$ is small and hence $\log(1+e^{-|\lambda_0^j|}) \approx 0$. Otherwise, we can approximate $\log(1+e^{|\lambda_0^j|})\approx |\lambda_0^j|$. The term $\log(1-p_{e,j})$ and $B=\sum_{j=0}^{N-1} \log(1-p_{e,j})$ can be pre-computed offline and can be used in the metric computation. 

Note that all the terms in (\ref{eq:pac_metric02}) are negative 
and so are the metric values. To save one bit per metric in the storage, we can discard the bit representing the always negative sign from the values. In this case we need to modify the comparisons in the algorithms accordingly. 


\begin{figure}
    \centering
    \includegraphics[width=0.7\columnwidth]{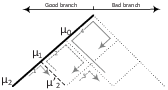}
    \caption{Bottom-up backtracking} 
    \label{fig:bu_backtracking}
    \vspace{-5pt}
\end{figure}
\begin{figure}
    \centering
    \includegraphics[width=0.7\columnwidth]{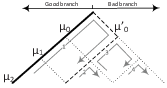}
    \caption{Top-down backtracking} 
    \label{fig:td_backtracking}
    \vspace{-5pt}
\end{figure}

\begin{algorithm}[t]
\small
\caption{PAC Fano Decoding}
\label{alg:fano_coding}
\DontPrintSemicolon
\SetKwInOut{Input}{input}
\SetKwInOut{Output}{output}
\SetKwRepeat{Repeat}{do}{while} 
\Input{Channel LLRs $\mathbf{\lambda_n^{0,N-1}}$, , $N$, $K$, $\mathcal{B}$, $\mathbf{p_{e}}$, $\mathbf{g}$, $\Delta$, $i_{bu}$}
\Output{Information bits $\mathbf{\hat{d}}$}
    $\mathcal{CS} \gets$ generateCS($\mathcal{B}$) \tcp*{Critical set \cite{rowshan}}
    cState[1,...,$|\mathbf{g}|-1$] $\gets$ \{0\} \tcp*{Current state}
    currState[0:$K$-1][1:$|\mathbf{g}|\!-\!1$] $\gets$ \{0\}\;
    [$i$, $j$, $T$, $\mathbf{\lambda}$, $\mathbf{\delta}$, $\mathbf{\beta}$, $b_{-1}$, $\alpha_q$] $\gets$ [0, 0, 0, \{0\}, \{0\}, \{0\}, $B$, 1]\;
    [onMainPath, isBackTracking, toDiverge, biasUpdated] $\gets$ [{\em True}, {\em False}, {\em False}, {\em False}]\;
    \While{$i < N$}{
        $\lambda_0^i\gets $ updateLLRs($i$, $\mathbf{\lambda}$, $\mathbf{\beta}$)\tcp*{\footnotesize like SCD}
        \uIf{$i \notin \mathcal{B}$}{
            [$\hat{u}_i$, cState] $\gets $ conv1bTrans($0$, cState, $\mathbf{g}$)\tcp*{\footnotesize  Alg. \ref{alg:pac_encoding}}
            
            $\mu_i \gets$ $\mu_{i-1}+$ $m$($\lambda_0^i$, $\hat{u}_i$)$-\alpha_q\cdot \log(1-p_{e,j})$\; 
            $\mathbf{\beta}\gets$ updatePartialSums($i, \hat{u}_i$, $\mathbf{\beta}$)\tcp*{\footnotesize like SCD}
            $i \gets i+1$\;
        }
        \Else{
            [$\hat{u}^{(0)}$, cState$^{(0)}$]$\gets$conv1bTrans($0$, cState, $\mathbf{g}$)\;
            [$\hat{u}^{(1)}$, cState$^{(1)}$]$\gets$conv1bTrans($1$, cState, $\mathbf{g}$)\;
            $\mu^{(0)} \gets$ $\mu_{i-1}+$ $m$($\lambda_0^i$, $\hat{u}^{(0)}$)$-\alpha_q\cdot \log(1-p_{e,j})$\;            
            $\mu^{(1)} \gets$ $\mu_{i-1}+$ $m$($\lambda_0^i$, $\hat{u}^{(1)}$)$-\alpha_q\cdot \log(1-p_{e,j})$\;
            [$\mu_{max}$,$\hat{v}_{max}$]$\gets$ [$\mu^{(0)}$,0] if $\mu^{(0)}>\mu^{(1)}$, else [$\mu^{(1)}$,1]\;
            [$\mu_{min}$,$\hat{v}_{min}$]$\gets$ [$\mu^{(0)}$,0] if $\mu^{(0)}<\mu^{(1)}$, else [$\mu^{(1)}$,1]\;
            \If{{\em onMAINpath}=True {\em\bf and} {\em isBackTracking} = True}{
                \uIf{$\mu_{min}>T$ {\em\bf and} $\mathcal{CS}[j]=1$ {\em\bf and} $j<j_{end}$}{
                    [onMAINpath, $\delta^s_j$, $j_{stem}$] $\gets$ [{\em False}, 1, $j$]\;
                    $[\mathbf{\lambda^s}, \mathbf{\beta^s}] \gets [\mathbf{\lambda}, \mathbf{\beta}]$\;
                    
                }
                \ElseIf{$j = j_{end}$}{
                    isBackTracking = {\em False}\;
                    $T=\lfloor \frac{\mu_{end}}{\Delta}\rfloor \Delta$\tcp*{\footnotesize Updating threshold}
                }
            }
            \uIf{$\mu_{max} > T$}{
                \uIf{{\em toDiverge} = False}{
                    [$\hat{v}_i$, $\hat{u}_i$] $\gets$ [$\hat{v}_{max}$, $\hat{u}^{(\hat{v}_{max})}$]\;
                    \uIf{{\em onMAINpath}= True {\em\bf and} $\delta^s_j$ = 1}{
                        [$\mu_i$, $\mu_i^{\prime}$] $\gets$  [$\mu_{max}$, $\mu_{min}$]\;
                    }\Else{
                        [$\mu_i$, $\mu_i^{\prime}$] $\gets$  [$\mu_{max}$, $\mu^{\prime\prime}_{i}$]\;
                    }
                    $\delta_j$ $\gets$ 0\;
                }
                \Else{
                    [$\hat{v}_i$, $\hat{u}_i$] $\gets$ [$\hat{v}_{min}$, $\hat{u}^{(\hat{v}_{min})}$]\;
                    [$\mu_i$, $\mu_i^{\prime}$] $\gets$  [$\mu_{min}$, $\mu_{max}$]\;               
                    [$\delta_j$, toDiverge] $\gets$ [1, {\em False}]\;
                    
                }
                [currState[$j$],cState]$\gets$[cState,cState$^{(\hat{v}_i)}$]\;
                $\mathbf{\beta}\gets$ updatePartialSums($i, \hat{u}_{i}$, $\mathbf{\beta}$)\;
                [$i, j$] $\gets$ [$i+1, j+1$]\;
            }
            \Else{
                \uIf{{\em biasUpdated} = $False$ {\em\bf and} $i<i_{bu}$}{
                    $T=\lfloor \frac{\mu_{max}}{\Delta}\rfloor \Delta$\tcp*{\footnotesize Updating threshold}
                }
                \Else{
                    $<$Go to {\bf Algorithm \ref{alg:fano_coding2}}$>$\;
                    \setcounter{AlgoLine}{67}
                }
            }
        }

    }
    \KwRet ($\mathbf{\hat{d}} \gets$ extract($\mathbf{\hat{v}}$, $\mathcal{A}$)) \tcp*{Dropping 0s}
\end{algorithm}
\normalsize

\begin{algorithm}
\caption{Lines 46-67 in Algorithm \label{alg:fano_coding2}
\ref{alg:fano_coding}}
\DontPrintSemicolon
\setcounter{AlgoLine}{45}
                    \If{{\em biasUpdated} =  False {\em\bf and} $i=i_{bu}$}{
                        \If{$\mu_{max} <$ B}{
                            $\alpha_q = \lceil\frac{\mu_{max}}{B\cdot\Delta_q}\rceil\Delta_q$\;
                            biasUpdated = {\em True}\;
                            \For{$k$ $\gets$ 0 \KwTo $j$}{
                                    $\mu^{\prime}_{\mathcal{B}[k]} = \mu^{\prime}_{\mathcal{B}[k]} + (\alpha_q-1)\cdot B^c_{\mathcal{B}[k]}$\;
                            }
                        $\mu_{\mathcal{B}[0]-1} = \mu_{\mathcal{B}[0]-1} + (\alpha_q-1)\cdot B^c_{\mathcal{B}[0]-1}$\;
                        $\mu_{\mathcal{B}[j]-1} = \mu_{\mathcal{B}[j]-1} + (\alpha_q-1)\cdot B^c_{\mathcal{B}[j]-1}$
                        }
                    }
                    currState[$j$] $\gets$ cState\;
                    \uIf{{\em onMAINpath} = False}{
                        \If{$\mu^{\prime\prime}_{\mathcal{B}[j_{stem}]} < \mu_{max}$}{
                            $\mu^{\prime\prime}_{\mathcal{B}[j_{stem}]} \gets \mu_{max}$\;
                        }
                    }
                    \Else{
                        [$j_{end}$, $\mu_{end}$] $\gets$ [$j$, $\mu_{max}$]\;
                        [frmMAINpath, isBackTracking] $\gets$ [{\em True}, {\em True}]\;
                    }
                    [T, $j^\prime$, toDiverge] $\gets$ moveBack($\mu^{\prime}_{0,i}$, $j$, $T$, $\delta_{0,j}$, $\mathbf{\hat{u}}$, $\mathbf{\mathcal{CS}}$, frmMAINpath)\tcp*{$\mu^{\prime}_{0,i}=\mu^{\prime}_0,\mu^{\prime}_1,...,\mu^{\prime}_i$}

                    \uIf{{\em toDiverge} = False {\em\bf and}  ($j^\prime = j_{stem}$ {\em\bf or} $j^\prime = j$)}{
                        onMAINpath = {\em True}\;
                    }\Else{
                        onMAINpath = {\em False}\;
                    }
                    [$i$, $j$, frmMAINpath] $\gets$ [$\mathcal{B}[j^\prime]$, $j^\prime$, {\em False}]\;                     cState $\gets$ currState[$j$]\;
\end{algorithm}

\begin{algorithm}
\caption{moveBack: Checking the previous examined nodes for moving backward}
\label{alg:move_back}
\DontPrintSemicolon
\SetKwInOut{Input}{input}
\SetKwInOut{Output}{output}
\SetKwRepeat{Repeat}{do}{while} 
\Input{the channel output $\mathbf{\mu^{\prime}}$, $j$, $T$, $\delta_{0,j}$, $\mathbf{\hat{u}}$, $\mathbf{\mathcal{CS}}$, frmMAINpath}
\Output{$T$, $j^\prime$, toDiverge, }
    isMovingBack $\gets$ False\;
    \While{True}{
        $j^\prime \gets j$\;
        \uIf{{\em frmMAINpath} = True}{\tcp*{\footnotesize Top-down move}
            \For{$k \gets 0$ \KwTo $j^\prime-1$}{ 
                \uIf{$\mu^{\prime}_{\mathcal{B}[k]}>T$ {\em\bf and} $\mathcal{CS}[k]=1$}{
                    $[j^\prime, j_{stem},  isMovingBack] \gets [k, k, True]$\;
                    $[\mathbf{\lambda^s}, \mathbf{\beta^s}] \gets [\mathbf{\lambda}, \mathbf{\beta}]$\;
                    break\;
                }
            }
            \uIf{$j^\prime = j$}{
                toDiverge $\gets False$\;
                \KwRet [T, $j$, toDiverge]\;
            }
        }
        \Else{ \tcp*{\footnotesize Bottom-up move}
            \For{$k \gets j^\prime-1$ \KwTo 0}{
                \uIf{$j_{stem} = k$}{
                    $j^\prime \gets k$\;
                    $[\mathbf{\lambda}, \mathbf{\beta}] \gets [\mathbf{\lambda^s}, \mathbf{\beta^s}]$\;
                    toDiverge $\gets False$\;
                    \KwRet [T, $j^{\prime}$, toDiverge]\;
                }
                \uIf{$\mu^{\prime}_{\mathcal{B}[k]}>T$ {\em\bf and} $\mathcal{CS}[k]=1$}{
                    \uIf{sum($\delta_{0,k}$) $\geq$ {\em maxDiversions}}{
                        continue\;
                    }
                    \uIf{$\delta_{k}$ = 1}{
                        [$j^\prime$, isMovingBack] $\gets [k, True]$\;
                        break\;
                    }
                }
            }
        }
        \uIf{{\em isMovingBack} = True}{
            $[i_{cur}, i_{start}] \gets [\mathcal{B}[j], \mathcal{B}[j^{\prime}]]$\;
            $[\mathbf{\lambda}, \mathbf{\beta}]\gets$ updateLLRsPSs($i_{start}, i_{cur}, \mathbf{\hat{u}}$, $\mathbf{\lambda}$, $\mathbf{\beta}$) \tcp*{\footnotesize Alg. \ref{alg:inter-LLR}}
            \uIf{$\delta_{j^\prime}$ = 0}{
                toDiverge $\gets True$\;
                \KwRet [$T$, $j^{\prime}$, toDiverge]\;
            }
            \ElseIf{$j^{\prime}$ = 0}{
                toDiverge $\gets False$\;
                \KwRet [$T$, $j^{\prime}$, toDiverge]\;
            }
        }
    }


\end{algorithm}

\subsection{Constrained Tree Search} \label{ssec:constrained_search}
The tree search algorithm may explore the paths on the tree that are unlikely to be correct. 
Unfortunately, the threshold T can only be used to prune a subset of these paths since a too tight threshold would also be likely to prune the correct path. Prior knowledge about error occurrence can be exploited in order to constrain the tree traversal. In the following, we propose several effective constraints resulting in a significant reduction in time complexity at a small performance degradation:
\subsubsection{Constraint on Number of Diversions from Best Path} 
\begin{figure}
    \centering
    \vspace{-5pt}
    \includegraphics[width=.75\columnwidth]{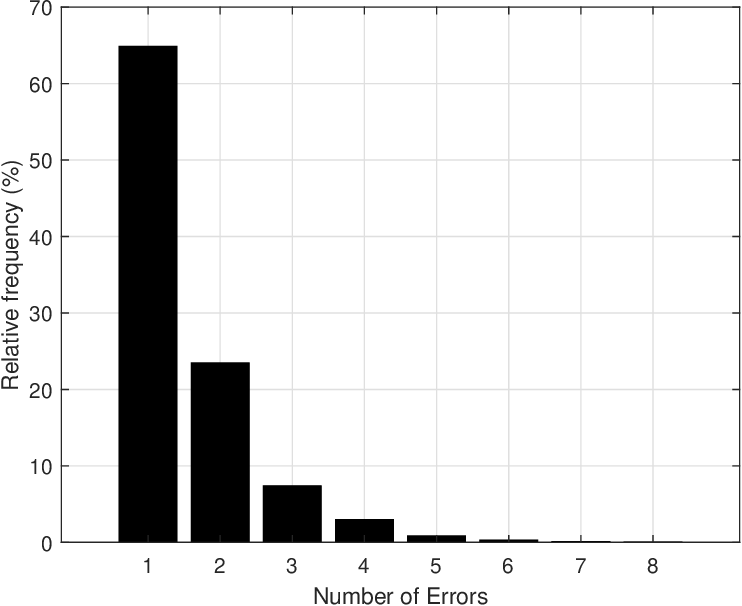}
    \vspace{-5pt}
    \caption{Distribution (in \%) of the number of error occurrence, extracted from 4000 decoding failures of PAC(128,64) with RM-profile at $E_b/N_0=2.5$ dB} 
    \label{fig:err_freq}
\end{figure}

By using a genie that corrects the error occurrence due to channel noise, we can observe that less than 1\% of the frame errors are due to more than $b=5$ bit-errors caused by the channel noise. Fig. \ref{fig:err_freq} shows the relative frequency of error occurrence for different numbers of bit-errors. 
With this knowledge, we can avoid exploring the paths that diverge from the SC path at more than 5 bit-positions. 
If we can afford a degradation of error correction performance, we can reduce the maximum number of diversions while exploring the alternative paths. This would limit the number of visited nodes. For the example shown in Fig. \ref{fig:err_freq}, this number can be set to $b=3$ or 4 bit-positions in order to reduce the number of visited node and consequently the time complexity. We will show a result after applying this constraint in Section\,\ref{sec:results}. In algorithm \ref{alg:move_back}, lines 21-22 implement the constraint for the maximum diversions.

\subsubsection{Exploring a Subset of bad branches} The reliability of the bit-channels is known from methods such as density evolution \cite{trifonov3}. Hence, during backtracking, we do not need to extend the partial path to the bad branches connecting to the nodes representing high-reliability bit-channels even if they satisfy the threshold condition. 
Thus, we only explore the sub-paths that originate from bad branches of the low-reliability bit-channels. This might introduce a small error rate degradation (due to not exploring all the bad branches), but it reduces the time complexity significantly. To this end, we collect the indices of the low-reliability bit-channels in the critical set $\mathcal{CS}$ \cite{cui,rowshan} and in the backtracking procedure, we only compare the threshold with the metrics of bad branches that are listed in the critical set. Lines 6 and 20 in Algorithm \ref{alg:move_back} enforce this constraint in top-down and bottom-up schemes (discussed in the next section), respectively.

Additionally, the constraint can be set to stop decoding and declaring decoding failure when the number of steps or clock iterations exceeds some limit or the path metric drops below a certain value. This could avoid cases with excessive run-time due to visiting a huge number of nodes. 
Also, we can stop decoding when the path metric drops below a certain value, since in this case, the decoder either fails  correcting the error(s) 
or it may lead to a long decoding delay due to visiting a huge number of nodes in order to find the correct path.

\subsection{Direction of Backtracking Traversal}\label{ssec:search-strategy}
Considering the properties of PAC codes which are mainly inherited from polar codes, we can devise different strategies that help to reduce the total number of nodes to visit during backtracking. 
When a decision error occurs during forward tree traversal, this error is propagated to the subsequent bits due to the sequential nature of decoding.  
In the conventional Fano decoding, backtracking starts from the latest decoded bit in a depth-first bottom-up direction, step by step as shown in Fig. \ref{fig:bu_backtracking}. For example, in a code with 3 bits, in the first backtracking iteration shown by 1 in Fig. \ref{fig:bu_backtracking}, the 3rd bit diverges from the SC path, i.e.,  $u_0-u_1-\bar{u}_2$. In the 2nd backtracking iteration, the 2nd bit  diverges only, i.e., $u_0-\bar{u}_1-u_2$. Then the 2nd and 3rd bits diverge together, i.e.,  $u_0-\bar{u}_1-\bar{u}_2$. This process continues towards the top of the tree until (in the worst case) all the combinations of 1-bit, 2-bit, and 3-bit diversions are explored, 
assuming the threshold condition is satisfied by all the branches.  
However, as our observations show, the probability that the first error due to channel noise has occurred at one of the  first decoded bits is higher. Further, there is no point in correcting the error that occurred due to error propagation.  Thus, backtracking in a top-down fashion as shown in Fig. \ref{fig:td_backtracking} is more consistent with the location of the first error and the subsequent propagated errors. 

The top-down backtracking can only be performed on the bad branches that originate from the SC path as a reference path. The rest of the backtracking iterations follows the bottom-up fashion. 
Note that a good branch is determined as a local branch with a higher likelihood among two branches emerging from a parent node. Thus, a good branch could form a non-SC path any where on the decoding tree. However, the SC path is distinguished by following the good branches at all the decoding steps from the root to the leaf of the tree. This SC path is shown by the bold line in Fig. \ref{fig:bu_backtracking} and Fig. \ref{fig:td_backtracking}.

Choosing a bad branch in the backtracking is called a {\em diversion} 
and its corresponding metric is denoted by a prime symbol, i.e., $\mu^\prime$, in Fig. \ref{fig:bu_backtracking} and Fig. \ref{fig:td_backtracking}. This diversion is equivalent to flipping a bit/bits \cite{afisiadis} from the SC path in the SC decoding. In Algorithm \ref{alg:move_back}, lines 5-12 and 14-25 implement the top-down and the bottom-up traversals, respectively.

\subsection{Threshold Update Strategy} When the channel noise has a high impact on the decision LLRs of low-reliability bits, as discussed in Section \ref{ssec:metric}, the best path metric $\mu$ drops significantly over a burst of low-reliability bit-channels such that $\mu \ll T$. On the other hand, at every iteration of backtracking (i.e., exploring all the potential sub-paths branching off from the current path), the threshold is reduced by $\Delta$. Thus, several backtracking iterations are required to satisfy $\mu>T - m\Delta$ for $m>1$ ($m$ is the number of backtracking iterations). If we skip the $m-1$ iterations and just we perform one iteration and then update the threshold at once using $T=\lfloor \frac{\mu}{\Delta}\rfloor \Delta$ to satisfy the condition $\mu>T$, we can proceed with the decoding of the  current best path and avoid extra delay. 
There is a possibility that the correct path is not the most likely path and the decoder could find another path in one of the backtracking iterations that we are going to skip. However, our observation shows that the degradation due to skipping $m-1$ backtracking iterations is about 0.05 dB at the high SNR regime.  The lines 24-26 in Algorithm \ref{alg:fano_coding} shows the implementation of this strategy. 

\begin{figure}
    \centering
    \includegraphics[width=1\columnwidth]{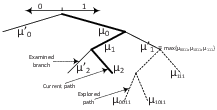}
    \caption{Updating the Metric of Explored  Branches} 
    \label{fig:metric_updating}
    \vspace{-5pt}
\end{figure}

\subsection{Updating Expected Metrics of Explored Paths}
During backtracking, the sub-paths originated from the current best path through bad branches are partially explored. The exploration of the same paths (possibly with longer length) might be repeated later as we proceed with the decoding. Our aim is to benefit from the time spent to explore the sub-paths.  
By updating the path metric, $\mu_j^{\prime}$, at the bad branch originated from the current best  path, as shown in Fig. \ref{fig:metric_updating}, with the actual result of the exploration rather than the expected path metric, we may avoid re-exploring these paths in the  next cycle(s) of backtracking. Since many sub-paths might originate from the same branch, we update $\mu_j^{\prime}$ with the largest metric obtained among sub-paths. This process is  performed in lines 55-57 of Algorithm \ref{alg:fano_coding2}. Here, we use $\mu^{\prime\prime}$ instead for temporarily storing the actual path metric of first sub-path explored and then comparing it with the actual metric of any new sub-path explored later. Then $\mu^{\prime}$ is updated in line 33 of Algorithm\,\ref{alg:fano_coding}. 
Note that by employing the adaptive heuristic metric, the effect of this updating becomes insignificant.



\section{NUMERICAL RESULTS} \label{sec:results}
In this Section, the error correction performance and the complexity of different tree search algorithms with different setups, using the previously discussed tree search complexity-reduction ideas and adaptive metric, are analyzed. 

To obtain the numerical results in this Section, we use different rate-profiles such as Reed-Muller (RM), density evolution with Gaussian approximation, and the polarization weight (PW) \cite{liu} with minimum row-weights eliminated.  Fig.\,\ref{fig:rate-profile} illustrates the aforementioned rate-profiles.  Here, we briefly revise the RM-profile and the modified PW-profile.
\subsubsection{Reed-Muller (RM) Rate-profile}\label{sssec:rm-polar} 
\begin{figure}[t]
    \centering
    \includegraphics[width=1\columnwidth]{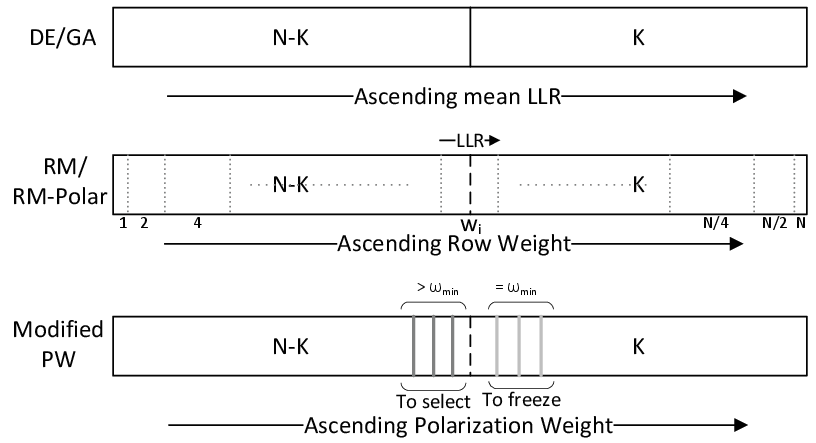}
    \caption{Rate-profile Schemes}
    \label{fig:rate-profile}
    \vspace{-8pt}
\end{figure}

The bit-channels for information bits are selected according to the row-weights ($w_i=wt(g_N^i)$ where $g_N^i$ is the $i$-th row) of $G_N$. When the candidate bit-channels with the smallest row-weight is more than need, the more reliable ones are selected. In this case, the rate-profile is called {\em RM-polar} \cite{li}\footnote{\color{blue}Note that the RM-Polar rate profile $\mathcal{A}^{\text{RM-Polar}}=\mathcal{A}^{\text{RM}}\cup\mathcal{A}^{\text{Polar}}$ suggested in Section \ref{sssec:rm-polar} is quite different from \cite{li} since $K'=\sum_{j=0}^{r'}{n \choose j}$ channel indices forming $\mathcal{A}^{\text{RM}}$ are chosen based on the row-weight rule, among $\mathbf{G}_N$-rows with weight larger than $d_{min}$, and the rest, that is, $K-K'\leq{n\choose r'+1}$ among the most reliable channels corresponding to rows with weight $d_{min}$, as demonstrated in Fig. \ref{fig:rate-profile} (middle). See the method {\em rmPolar\_build\_mask} in rate\_profile.py on github.}. In this work, the reliability measure is the mean LLR obtained from density evolution with Gaussian approximation (DEGA).
\subsubsection{A Modified Polarization Weight (PW) Rate-profile} In this method, the bit-channels for information bits are selected among the ones with the largest polarization weight ($W_i$), $W_i=\sum_{j=0}^{n-1}b_j\cdot 2^{j\cdot\frac{1}{4}}$, where $i=b_{n-1}...b_{0}$ is the binary representation of $i$ \cite{liu}. 
In order to improve the distance property, we propose to freeze the selected bit-channels with minimum row-weight and replace them with the bit-channels with lower $W_i$, but larger $w_i$. 

In the simulations, we employ different generator polynomials (0o36, 0o133, 0o177, and 0o1563 in octal format) with  constraint lengths 5,7,7, and 10, respectively. 
The numerical results show that the difference among them in terms of FER is negligible in the low SNR regime and small in high SNRs. 

Finally, for the purpose of comparison in the figures, we use the dispersion bound \cite{polyanskiy} a.k.a. Polyanskiy-Poor-Verdu (PPV) bound or finite-length bound which is a Gaussian approximation on the block error probability of finite-length block codes.  
Additionally, we employ lower bound on ML performance as well. This bound is obtained under list decoding with $L=256$ by assuming that ML decoder would fail when $\mathbf{\hat{v}\neq v}$ but $\sum_{i=0}^{N-1} ||\hat{x}_i-y_i||<\sum_{i=0}^{N-1} ||x_i-y_i||$ where $\mathbf{\hat{x}=\hat{v}GP_n}$.

\subsection{Distance Spectrum}
\label{ssec:distance-prop}
As discussed in Section \ref{sec:PACs}, by convolutional pre-coding, we are no longer transmitting fixed known values, e.g., 0 frozen bits, over low-reliability (bad) synthetic channels, but random values generated by a linear combination of information bits. 
To analyze the impact of this difference on polar codes, we use the multilevel SCLD-based search method in \cite{zhang2} to enumerate the codewords with the minimum Hamming distance, $d_{min}$. We use the size of $L=2^{17}$ and in each iteration we introduce a one-bit error in the positions corresponding to the minimum row weight in $P_n$, when the all-zero codeword is transmitted and no noise is added. Re-encoding the candidate messages, remaining in the list at the end of decoding, shows that the number of codewords with the minimum Hamming weight $d_{min}=16$ is $A_{16}=94488$ for the polar code {\em P}$(128,64)$ constructed with RM-profile, whereas $A_{16}=3120$ for the PAC code {\em PAC}$(128,64)$ with the same rate profile. Furthermore, the second minimum distance for the polar code is 24 with $A_{24}=4465024$ while for the PAC code we observe $A_{18}=2696$, $A_{20}=95828$, $A_{22}=352311$ and $A_{24}=3065194$. Note that the minimum Hamming distance for {\em PAC}$(128,64)$ with PW \cite{liu} rate profile is $d_{min}=8$ with $A_{8}=256$ and $A_{12}=960$, hence the FER performance of PW-profile is inferior to RM-profile. Hence, PW-profile for PAC(128,64) is not considered. 

From the truncated union bound of the block error probability under ML decoding, $P_e^{ML} \approx A_{d_{min}} Q(\sqrt{2d_{min}R E_b/N_0})$ \cite{moon}, we can conclude that given the same $d_{min}$ and decoder, the code with smaller $A_{d_{min}}$ should perform better. 
In \cite{li2}, the authors show that a properly designed upper-triangular pre-transformation matrix for polar codes can reduce $A_{d_{min}}$ of the concatenated code. Note that the convolutional pre-transform in PAC codes has an upper-triangular Toeplitz  matrix.

\subsection{List Decoding}\label{ssec:results-list}
\begin{figure}[ht]
    \centering
    \includegraphics[width=1\columnwidth]{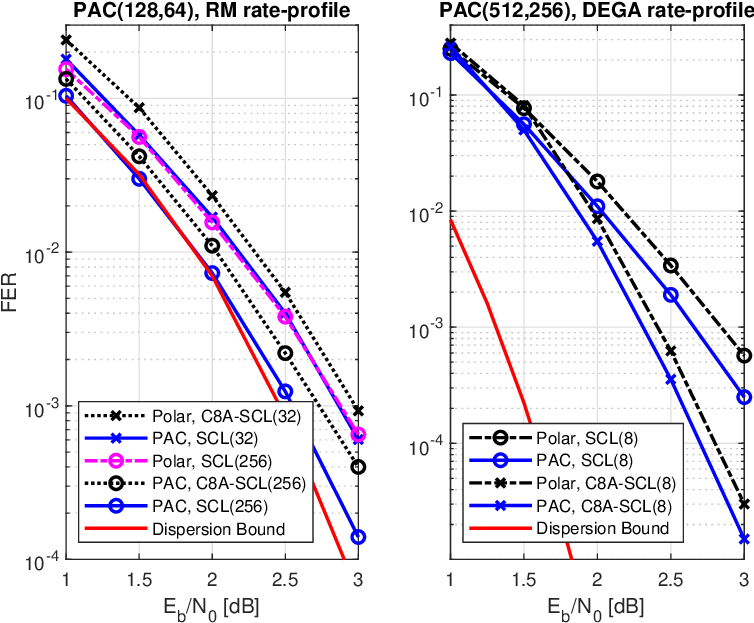} 
    \caption{Performance of PAC codes under list decoding}
    \label{fig:fer_list}
    \vspace{-10pt}
\end{figure}

The list decoding of PAC codes over binary-input  additive white Gaussian noise (BIAWGN) channels with BPSK modulation is simulated. The constraint length and the coefficients of the generator polynomial for the convolutional code are 7 ($m=6$) and 0o133, respectively. For PAC(128,64), the rate-profile is formed by the Reed-Muller (RM) construction \cite{li} with dSNR=3.5. In the list decoding, different list sizes are employed and the performance is compared with the performance of the P(128,64) polar code and finite-length bound \cite{polyanskiy} as shown in Fig. \ref{fig:fer_list}. The performance of the RM-profile and the modified PW-profile are  identical as the resulted rate-profiles are identical. A serial concatenation of CRC with relatively short codes such as PAC(128,64) does not  improve the error correction performance due to a significant rate loss and negative impact on the distance properties (e.g. in the case of PAC(128,64), the minimum Hamming distance drops to $d_{min}=8$). However, an 8-bit CRC with a generator polynomial with coefficients 0xA6  improves the performance of PAC(512,256) in the high SNR regime significantly as shown in Fig. \ref{fig:fer_list}. The notation CxA-SCL used in Fig. \ref{fig:fer_list} is defined as CRC-aided SCL decoding with x-bit CRC and $L$ in SCL($L$) is the list size. The rate-profile for this code is formed by density evolution with Gaussian approximation (DEGA) \cite{trifonov3} with dSNR=2. One can observe that as the block-length increases, the performance of PAC codes under list decoding cannot compete with that of polar codes under CRC-aided list decoding and we need to add CRC bits as the outer code to detect the correct path in the list decoding. 

\subsection{Fano Decoding}\label{ssec:results-fano}
\begin{figure}[t]
    \centering
    \includegraphics[width=1\columnwidth]{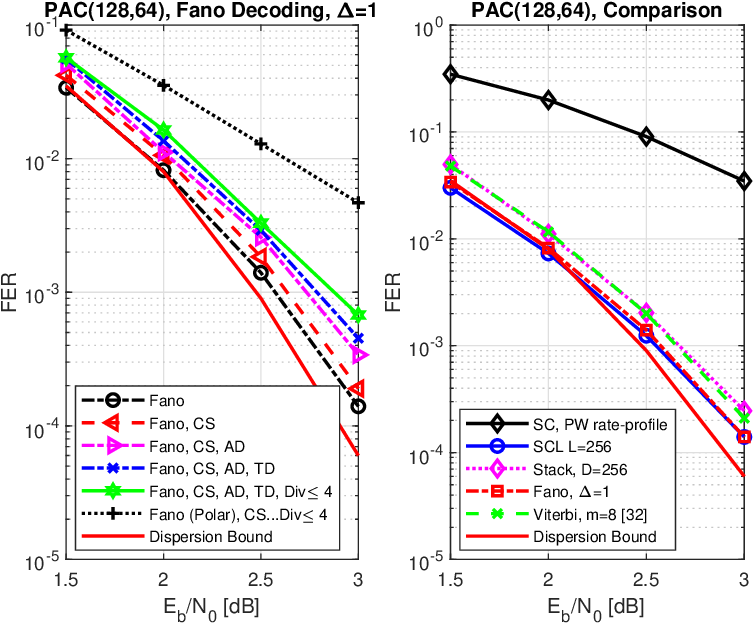}
    \caption{Performance of PAC codes with RM rate-profile under Fano decoding  with constrained search (CS), adaptive metric (AD), top-down tree traversal (TD), and a limited number of diversions (Div.) in comparison with other decoding schemes SC, SCL, stack, and Viterbi. Also showing performance of polar codes under Fano decoding  "Fano (Polar)".}
    \label{fig:fer_fano}
    \vspace{-10pt}
\end{figure}

\begin{figure}[t]
    \centering
    \includegraphics[width=1\columnwidth]{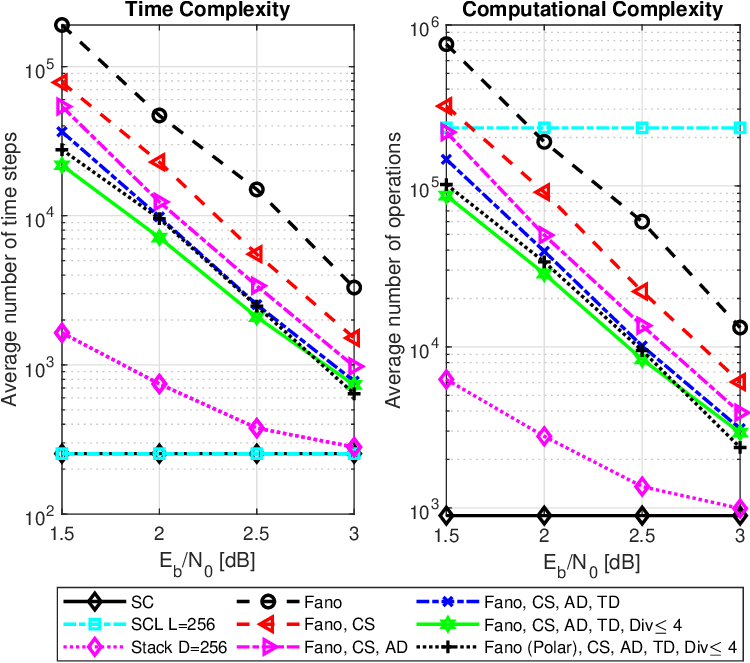}
    \caption{Time and computational Complexity.}
    \label{fig:cplx_fano}
    \vspace{-10pt}
\end{figure}

The Fano decoding algorithm provides a performance near the dispersion bound, but as a variable-complexity decoding scheme, its average time complexity is extremely high. The Fano decoding of PAC(128,64) with RM-profile 
over BIAWGN channel is simulated. Similar to list decoding, the constraint length and the coefficients of the generator polynomial for convolutional codes are 7 ($m=6$) and 0o133, respectively. The non-optimized design-SNR for obtaining the pre-computed bias term is 4 dB. By applying the ideas  introduced in Section \ref{sssec:fano-decoding}, such as adaptive metric (AD), top-down (TD) search strategy and imposing constraint on the number of diversions (Div) from SC path, it is observed in Fig. \ref{fig:fer_fano} (left) and Fig. \ref{fig:cplx_fano} (left) that while the average time complexity drops significantly by 50\% to 80\%, depending on the  techniques employed, the degradation in error correction performance is not high. Since the curves in Fig. \ref{fig:cplx_fano} are almost straight in semi-logarithm scale, the complexity gains are preserved at high SNR regimes as well. 
Fig. \ref{fig:cplx_fano} (right) shows the computational complexity of Fano, stack, and list decoding under different parameters and techniques. The computational complexity is measured by the total number of operations per codeword (comparisons and additions) performed through the  factor graph in Fig. \ref{fig:factor_graph}.  As can be seen, the computational complexity of list decoding is significantly higher than Fano and stack decoding to achieve the same performance. 
Fig. \ref{fig:cplx_fano} (left)  shows the time complexity in terms of time steps, where  each time step is defined as the time required for processing the node(s) in one stage of the factor graph shown in Fig. \ref{fig:factor_graph}.
Although the time complexity of the list decoding is significantly lower than Fano and stack decoding (left),  we note that one time step in list decoding is longer than a time step in Fano decoding, due to the required sorting process. 

Note that stack decoding has a lower time and computational complexity than Fano decoding because it does not need to trace back on the tree and explore other paths to find a promising one if there is any. The partial paths (sorted with respect to the metric) and their associated intermediate  information are already available in the stack. Hence, stack decoding can save a significant amount of computations and time at the cost of a huge memory requirement. 
To compute the number of time steps (or clock cycles), we consider an architecture that is similar to that in \cite{leroux}. In this type of design, 
$2N-2$ time steps are required to decode a codeword \cite{leroux}. However, in Fano decoding, due to possible backtracking, 
the number of required time steps is typically significantly larger than $2N-2$. Here, we take the average time steps over a large number of  decoding iteration into account. 
For comparison, we also implemented Fano decoding for polar codes with RM-profile. Although, the average computational complexities of polar codes and PAC codes under Fano decoding are close, due to poor weight distribution of polar codes, PAC codes outperform polar codes. 

Another important observation in Fig. \ref{fig:fer_fano} (left) is that the performance gain of PAC codes over polar codes under Fano decoding is quite significant while the time and computational complexity of these two families of codes are close. However, this performance gain under list decoding as shown in Fig. \ref{fig:fer_list} is smaller. 
Additionally, one can observe from the comparison of the performance of PAC(128,64) under list, stack, and Fano decoding in Fig. \ref{fig:fer_fano} (right) that Fano decoding provides a similar performance as list decoding but outperforms the stack decoding, while it requires significantly less hardware resources than list decoding and stack decoding. As shown in Table \ref{table:hardware_res}, the memory required for paths, intermediate LLRs and partial sums,
which account for the majority of memory space, for list and stack decoding is $L$ and $D$ times that of Fano decoding. Note that in order to obtain a FER performance similar to Fano decoding, we need a very large list size $L$ or stack depth $D$ in the order of 128 or 256. This highlights the huge gap between Fano decoder and the other decoders in terms of hardware resources 

\begin{table}[h!]
\centering
\begin{tabular}{ |c|c|c|c| } 
\hline
 & Fano & Stack & List \\
\hline
\multicolumn{4}{|l|}{\em Memory Requirement [bits]} \\
\hline
Path memory, $\mathbf{u}$  & $N$ & $DN$ & $LN$ \\ 
\hline
Intermediate LLRs, $\mathbf{\lambda}$  & $(N\!-\!1)Q_1$ & $D(N\!-\!1)Q_1$ & $L(N\!-\!1)Q_1$ \\
\hline
Partial Sums, $\mathbf{\beta}$  & $N\!-\!1$ & $D(N\!-\!1)$ & $L(N\!-\!1)$ \\
\hline
Path Metric, $M$ & $2(N\!-\!K)Q_2$ & $DQ_2$ & $LQ_2$ \\
\hline
Current State & $K\cdot m$ & $D\cdot m$ & $L\cdot m$ \\
\hline
Critical Set flag, $CS$ & $N$ & 0 & 0 \\
\hline
Diversion flag, $\delta$ & $N$ & 0 & 0 \\
\hline
Error probability, $p_e$ & $NQ_3$ & $NQ_3$ & 0\\
\hline
\multicolumn{4}{|l|}{\em Computing Resources} \\
\hline
Processing Elements & $P$ & $P$ & $LP$\\
\hline
Comparison & A comparator & $D$-sorter & $2L$-sorter\\
\hline
\end{tabular}
\caption{Comparison of hardware resources}
\label{table:hardware_res}
\end{table}

The parameters $P$, and $Q_i$ for $i=1,2,3$ denote the number of processing elements (PE) \cite{leroux} and the number of quantization bits,  respectively.

Finally, Viterbi algorithm (VA) \cite{rowshan-lva} with similar hardware resources as list decoding (except the $2L$-value sorter, replaced by a 2-value comparator)  provides a close performance to Fano and list decoders. 


\section{CONCLUSION} 
In this paper, we investigate the implementation of list decoding and Fano decoding for PAC codes. 
Under list decoding, there is a significant performance gap between polar codes and PAC codes. However, this gap between polar and PAC codes is reduced when employing another layer of concatenation, such as CRC or parity check (PC) bits. Also, the results show that a large list size $L$ or stack depth $D$  of $256$ under list and stack decoding, respectively, is needed to approach the performance of PAC codes under Fano decoding. 

Fano decoding has a large average time complexity but a small computational complexity relative to list decoding. For mitigating the time complexity, we propose several techniques and strategies including adaptive path metric and a heuristic to estimate a metric for the continuation of the partial paths, search constraints, and a combination of top-down and bottom-up search strategies. This strategies reduce the computational complexity as well. 
Also, to overcome the difficulty of obtaining the intermediate LLRs and partial sums during backtracking, we propose an algorithm to compute these intermediate information (LLRs and partial sums) efficiently without using extra memory to store them or any need to restart the decoding process. The numerical results show that by using these techniques, the average time complexity drops by 50\% to 80\% at the cost of a relatively small performance degradation. The adaptive heuristic metric and the search strategies proposed in this paper can be used in polar coding as well. Although the time complexity of the Fano Decoding is variable and high, the software Fano decoder is significantly faster than software list decoder with large list size without using parallelism.  

Due to need for  backtracking in Fano decoding, as the code-length increases, the frequency of backtracking through the decoding increases prohibitively. Hence, we conclude that the Fano decoding can be used for short codes with medium to low code rates. 

Overall, it appears that any proper pre-transformation such as convolutional transform \cite{arikan2}, moving parity check bits \cite{zhang}, dynamic frozen bits \cite{trifonov2}, use of CRC bits for error detection \cite{tal}, and a combination of them can improve the distance spectrum and results in an error correction performance gain. However, each pre-transformation may provide a different gain  depending on the rate-profile, block-length and code rate. 

\section*{ACKNOWLEDGMENT}
The authors are grateful to the anonymous reviewers for their useful comments and
suggestions which improved the clarity and the inclusiveness of the paper.

\end{document}